\documentclass[onecolumn,preprintnumbers,a4paper]{revtex4}
\usepackage{subfigure}     
\usepackage{graphicx}
\usepackage{amsmath, amssymb}
\usepackage{braket}
\usepackage{hyperref}
\usepackage{tabularx}

\begin{document}

\title{Cosmic searches for Lorentz invariance violation}

\author{Carlos P\'erez de los Heros$^{1}$ and Tomislav Terzi\'{c}$^{2}$}

\affiliation{$^{(1)}$Department of Physics and Astronomy, Uppsala University, Uppsala, Sweden.\\ 
$^{(2)}$Department of Physics, University of Rijeka, Rijeka, Croatia.\label{RIJ}
}

\begin{abstract}
 Cosmic messengers (gamma rays, cosmic rays, neutrinos and gravitational waves) provide a powerful complementary way to search for Lorentz invariance violating effects to laboratory-based experiments. The long baselines and high energies involved make Cherenkov telescopes, air-shower arrays, neutrino telescopes and gravitational wave detectors unique tools to probe the expected tiny effects that the breaking of Lorentz invariance would cause in the propagation of these messengers, in comparison with the standard scenario. In this chapter we explain the expected effects that the mentioned detectors can measure and summarize current results of searches for Lorentz violation.
\end{abstract}

\date{\today}
\maketitle

\section{Introduction}
\label{sec:intro}

The invariance of physical laws under Lorentz transformations is a fundamental requirement that has successfully guided the development of our current theory describing the basic constituents of matter and their interactions, the Standard Model of Particle Physics. Its meaning, that experimental results are independent of the orientation and velocity of the frame of reference, seems also a reasonable criterion for any physical theory. Therefore any deviation from this, in principle, hard-wired requirement of the Standard Model would indicate the emergence of new physics at some given energy scale where the theory would break down, see Chapter~1. We do know that the Standard Model is not the ultimate description of nature, even if only because it does not describe Gravity alongside the other fundamental forces. But there are other reasons why the Standard Model needs to be extended, like the existence of neutrino masses, the strong CP problem or the uncomfortably, to some, number of free parameters in the model that  an extended theory could account for. Besides, astrophysical observations, like the existence of dark matter or dark energy, need ultimately to be explained from a fundamental physics point of view and, whatever that explanation might turn out to be, it can not come from within the Standard Model. Not all the extensions of the Standard Model require or predict Lorentz invariance violation (LIV) but, contrarily, the observation of LIV effects would undoubtedly point to physics beyond the Standard Model.

 Breaking of Lorentz invariance is expected in models of quantum gravity with a minimal fundamental length that must be an invariant for all reference frames. This leads to a breakdown of Lorentz invariance at energies near the Planck scale. A consequence is that ultrarelativistic particles (small Compton wavelengths) propagating in vacuum between two points (i.e., source and detector) will perceive this quantum structure of space-time and cover the distance in a different time compared to propagation through classical, continuous, space-time. But LIV is not necessarily  connected to the existence of a discrete space-time. Lorentz invariance breaking operators can be added in an effective field theory formulated in standard space-time. From an experimental point of view, searches for LIV tend to be carried out in a model independent way, just parameterizing the effect of LIV in power expansions of the free particle Lagrangian as we discuss below, and setting limits on the parameters from the lack of observation of any LIV effect. In particular the 
 effect of LIV in the propagation of a particle can be parameterized quite generically as a modification of the dispersion relation, which acquires corrections at a given high-energy scale denoted below by $E_\mathrm{QG}$. The usual starting point of experimental searches for Lorentz invariance violation (LIV) is therefore the modified dispersion relation
\begin{equation}
    E_{\mathrm{i}}^2 = m_{\mathrm{i}}^2 c^4 + p_{\mathrm{i}}^2 c^2 \left[ 1 + \sum_{n=1}^{\infty} \eta^{\mathrm{(i)}}_n \left( \frac{p_{\mathrm{i}}c}{E_{\mathrm{QG,}n}^{\mathrm{(i)}}} \right)^n \right],\label{eq:generalMDR}
\end{equation}
where the index i represents the particle type (e.g. photon, electron, proton, neutrino, pion, etc.), $c$ is the standard special-relativistic invariant speed of light, $E_{\mathrm{i}}$, $p_{\mathrm{i}}$, and $m_{\mathrm{i}}$ are the particle's energy, momentum, and rest mass, respectively (n.b. $m_{\gamma} = 0$). The series expansion in the square brackets represents a modification to the standard (no LIV) dispersion relation, presumably caused by quantum gravity. Each term is characterised by $E_{\mathrm{QG,}n}^{\mathrm{(i)}}$, the energy level at which that particular contribution becomes relevant. Typically, these are expected to be on the scale of the Planck energy, but it does not necessarily need to be so. The 
parameter $\eta^{\mathrm{(i)}}_n$ is either $+1$ or $-1$. It determines the sign of the modifying contribution, corresponding to superluminal or subluminal effects respectively.

It is important to stress that the modified dispersion relation is not a consequence of any particular model of quantum gravity. Equation~(\ref{eq:generalMDR}) is a rather simple parameterization, a model used to test LIV. A possible detection of an effect will not necessarily prove any particular theory of quantum gravity. However, it will trigger research to determine the cause and nature of the violation. 
LIV can be formally incorporated in extensions of the Standard Model in the form of effective theories that explicitly include LIV terms in a consistent manner. The most popular of these approaches is the Standard Model Extension (SME)~\cite{Colladay:1996iz,Colladay:1998fq}. The SME is a renormalizable quantum field theory that inherits the $SU(3) \times SU(2) \times U(1)$ group structure of the SM but extends its Hamiltonian to include general Lorentz violating terms. These terms introduce new effects like energy-dependent velocity of masless particles, modified interactions, modified neutrino oscillations, a possible direction dependence of the photon polarisation, photon instability, CPT violation or the possibility of $\nu - {\bar{\nu}}$ mixing or neutrino bremsstrahlung, giving rise to a rich phenomenology~\cite{Kostelecky:2003cr,Hooper:2005jp,Kostelecky:2011gq,Diaz:2014yva}.

Although we will not go deep in discussing the underlying theories in this chapter (we refer the interested reader to chapters 1 and 2 for discussions on theory and phenomenology), it is worth mentioning that there are other scenarios than the SME in which the Lorentz symmetry is not strictly preserved as we know it, as in doubly special relativity (DSR) models~\cite{Amelino-Camelia:2002cqb,Amelino-Camelia:2010lsq}. The underlying principles of the SME and DSR are fundamentally different. As mentioned above, the SME is an effective theory that adds Lorentz violating terms to the Standard Model. On the contrary, DSR, keeps the speed of light $c$ as an observer invariant and a limiting propagation speed for low-energy photons, but adds an additional observer invariant in the form of a fundamental energy $E_{\mathrm{QG}}$. Typically, $E_{\mathrm{QG}}$ is taken as the Planck energy but, in some models, it can be orders of magnitude below it. In order to  accommodate this new invariant, Lorentz symmetry needs to be deformed, which in turn leads to deformed expressions for conservation laws, resulting in rather different predictions on the propagation of particles. In addition, some phenomena expected in LIV are not allowed in DSR.  We will point these out as we go along.

 Although no experimental evidence of violation of Lorentz invariance has been found yet, experimental searches on many fronts are being carried out, from laboratory experiments~\cite{Adelberger:2009zz,Safronova:2017xyt,Ivanov:2019ouz,Zhang:2021sbx} to astrophysical probes, the focus of this chapter. In what follows we describe the effects to be expected from LIV using cosmic messengers as probes, the experimental techniques used and the current experimental limits obtained with Cherenkov telescope arrays, neutrino telescopes, air shower arrays, and gravitational wave detectors.

\section{Cosmic messengers as a probe of LIV}
\label{sec:signatures}

\subsection{Gamma rays}
\label{subsec:GRs}

Gamma radiation constitutes a significant portion of the electromagnetic spectrum. Astrophysicists classify every photon with energy above 100\,keV as a gamma ray, regardless of the process in which it was created. Considering that the most energetic photon detected up to date was a 1.4\,PeV ($1.4 \times 10^{15}$\,eV)~\cite{2021Natur.594...33C}, the gamma-ray band spans over more than 10 orders of magnitude in energy. 

Compared to some other cosmic messengers (e.g. neutrinos in Section~\ref{subsec:NUs}), gamma rays are relatively easily detected and, since they are electrically neutral, they propagate on straight lines (compared to cosmic rays. See Section~\ref{subsec:CRs}), meaning that their direction of arrival points towards their source. These characteristics, combined with their high energies, make gamma rays superb probes of the most energetic processes in the universe, particularly convenient for searches of LIV.

\subsubsection{Gamma ray detectors}
\label{sec:Gamma-rayDetectors}
Gamma rays do not penetrate the atmosphere. Therefore, gamma-ray detectors either need to be placed on satellites in orbits above the Earth's atmosphere, or the atmosphere needs to be used as a target in the detection process. 
In either case, the underlying process is fundamentally the same, though at different scales. When a gamma ray enters the atmosphere, it interacts with atomic nuclei in the air. It is absorbed and an electron--positron pair is created. Both of them lose energy through interactions with other nuclei, emitting additional gamma rays in the process through \textit{bremsstrahlung}. These secondary gamma rays are also absorbed, and new e$^{-}$--e$^{+}$ pairs are created. A cascade of particles, known as extensive air shower (EAS), develops for as long as bremsstrahlung is the  preferred way for electrons to lose energy. Charged particles in the cascade can be energetic enough to propagate faster than light through the medium, instantaneously polarising molecules in the air as they pass through. As the molecules coherently depolarise, they emit flashes of so-called Cherenkov radiation, which peaks in the  ultraviolet band. Photomultiplier tubes in Cherenkov detectors record the Cherenkov light emitted as a final consequence of a gamma ray penetrating the atmosphere. Shower images are usually ellipse-shaped, with the longer axis pointing towards the position of the gamma-ray source in the camera, while the image size is connected to the energy of the primary particle.  Using imaging techniques, energies and directions of primary gamma rays are estimated. 

Showers induced by gamma rays contain predominantly electrons, positrons and photons, and their developments is governed through electromagnetic interaction. For that reason, they are often referred to as electromagnetic showers. Cosmic rays also induce extensive air showers, with several subtle but important differences. Unlike gamma rays, primary cosmic rays survive interactions with atomic nuclei in the medium. More importantly, in addition to the electromagnetic, the weak and strong interactions play important roles in the shower development. As a consequence, the so-called \textit{hadronic} showers are composed of secondary baryons, mesons (such as kaons or pions), and leptons (electrons, muons, neutrinos, etc.).  Hadronic showers often contain electromagnetic sub-showers. 

Hadronic showers are less homogeneous than gamma showers, which can be used to classify the type of the primary particle. As far as gamma-ray observations are concerned, cosmic rays constitute background, and their rate is in most cases much higher. That means that the region in the detector where the signal is expected will be contaminated with background events. The background flux is usually estimated from another region in the detector where no sources of gamma rays are expected. It is important to note that while the number of  background events in the signal region can be estimated, as of yet there is no way of determining which specific event belongs to the signal and which to the background. 
This is quite a drawback in research which is performed on single events, which is the case in some LIV studies.

The most important present-day Cherenkov telescopes are the Highe Energy Stereoscopic System 
H.E.S.S.~\cite{HESS:01, HESS:2015cyv, Holler:2015tca}, the Major Atmospheric Gamma Imaging Cherenkov (MAGIC)~\cite{MAGIC:01,         Aleksic:2014poa, MAGIC:2014zas} and the Very Energetic Radiation Imaging Telescope Array System (VERITAS)~\cite{VERITAS:01, Park:2015ysa}, while the next generation Cherenkov experiment, the Cherenkov Telescope Array (CTA)~\cite{CTA:01, CTAConsortium:2017dvg} is currently being constructed.
More details on the extensive air showers and Cherenkov effect one can find in~\cite{Spurio2015, DeAngelis2021}, while for a detailed review of data analysis techniques in Cherenkov telescopes we refer interested reader to~\cite{DAmico:2022psx}. 

We already mentioned that the shower size is correlated to the  energy of the primary particle. Therefore, the detector design will strongly depend on the targeted energy range. 
Starting with the lowest energies, satellite detectors are used to detect gamma rays with energies up to few hundred GeV. For example the Fermi Large Area Telescope (\textit{Fermi}-LAT)~\cite{Fermi-LAT:01, Fermi-LAT:2009ihh} is composed from two main parts. The first part consists of layers of conversion foils and tracking detectors, where a gamma ray is converted to an e$^{-}$--e$^{+}$ pair, and their trajectories are tracked to determine the direction of primary gamma ray. Showers are initiated in the calorimeter placed just below, where the energies of all shower constituents are summed to measure the energy of the primary particle. \textit{Fermi}-LAT is sensitive in the energy range 20\,MeV -- 300\,GeV. Lower energies can be accessed if Compton scattering is used in the detector instead of pair creation. Some future space detectors, e.g. enhanced.ASTROGAM (e-ASTROGAM)~\cite{e-ASTROGAM:2017pxr}, or the All-sky Medium Energy Gamma-ray Observatory (AMEGO)~\cite{AMEGO:01, AMEGO:2019gny}, will combine these processes to access lower energies, as well as measure the gamma-ray polarisation. 
Another deciding factor for detector design selection is the rate of gamma rays, which, as a general rule, falls off as a power-law of energy, requiring larger detector collection areas to access higher energies. The collection area of \textit{Fermi}-LAT of $\lesssim1$\,m$^2$ is too small to detect a relevant number of gamma rays with energies above $\sim300$\,GeV. In addition, the showers at these energies are too big to fit in the calorimeter, making gamma-ray energy reconstruction unfeasible. 

Cherenkov telescopes on the ground are, on the other hand, sensitive to gamma rays with energies from $\sim$10~GeV to $\gtrsim$100~TeV. Showers below the low energy threshold produce Cherenkov radiation too weak to be detected and can not be properly reconstructed with these instruments. 
As we go to higher energies, gamma rays become too rare, and air showers too extended to be detected with Cherenkov telescopes. However, charged particles within the showers reach the ground and can be detected by water Cherenkov detectors. Experiments such as the High Altitude Water Cherenkov Observatory (HAWC)~\cite{HAWC:01, Abeysekara:2017mjj} employ arrays of water containers with photomultipliers, which record flashes of Cherenkov light in the water, not unlike water neutrino detectors (see Section~\ref{subsec:NUs}). There are also hybrid detectors, as the Large High Altitude Air Shower Observatory (LHAASO)~\cite{LHAASO:01, LHAASO:2019qtb}, which combine several detection techniques to detect the highest energies and to suppress the background.
It is important to notice the significant overlap in energies covered by detectors of different types, which is essential for cross-instrumental calibration and collaboration.

While the gamma-ray detection technique is fundamentally the same in all these instruments, the implementations, and therefore the observation strategies, are quite different.
Instruments onboard the \textit{Fermi} satellite have a wide field of view. The \textit{Fermi}-LAT field of view covers almost 20\% of the sky, and mostly observes in the sky survey mode. It scans the entire sky in 3 hours, with any given point remaining in the field of view continuously for at least 30 minutes. Cherenkov telescopes, on the other hand, observe only small portions of the sky at the time, with fields of view of up to $\sim5\,\deg$. Therefore, they perform pointed observations, with usually only one source in the field of view. Considering that Cherenkov telescopes rely on detecting flashes of optical and UV light in the air, they cannot be used during the day, or in bad weather. Satellite detectors, do not suffer of such restrictions. Water Cherenkov detectors have a field of view comparable to \textit{Fermi}-LAT (e.g. HAWC instantaneously covers 15\% of the sky) and also perform sky surveys. However, unlike satellite-borne detectors or Cherenkov telescopes, they cannot be pointed nor repositioned. Water Cherenkov detectors also use photomultipliers, but these are enclosed in light-tight water tanks, so their duty cycle is close to 100\%, as in satellite detectors.

\subsubsection{Effects of Lorentz invariance violation}
\label{sec:LIVeffects}
Effects of modified photon dispersion relation can be most generally classified in two categories: i) modification of the propagation speed, and ii) modifications of photon interactions. 

\paragraph{\textbf{Energy dependent group velocity}}
Assuming that the group velocity still corresponds to the derivative of energy with respect to momentum, one can easily derive from Equation~(\ref{eq:generalMDR})

\begin{equation}\label{eq:photonvelocity}
    v_\gamma = \frac{\partial E_\gamma}{\partial p_\gamma} \simeq c \left[1+\sum_{n=1}^{\infty}\eta_n\frac{n+1}{2}\left(\frac{E}{E_{\mathrm{QG,}n}}\right)^{n}\right].
\end{equation}
Obviously, the modification introduced in the photon dispersion relation makes the modified photon group velocity energy dependent, and different from $c$. Depending on the value of $\eta_n$, the group velocity can be greater or less than the standard speed of light. These two behaviours are known as \textit{superluminal} ($\eta_n = +1$) and \textit{subluminal} ($\eta_n = -1$), respectively.

Assuming that the photon speed is not constant but that it depends on energy, two photons of different energies will have different times of flight (often abbreviated: ToF). Therefore, by comparing times of flight of photons of different energies, we can calculate the difference in their speeds. Based on the derivation by Jacob \& Piran in \cite{Jacob:2008bw}, the delay in the arrival times of two photons of different energies, emitted at the same time from the same source at redshift $z_{\mathrm{s}}$ is given by:

\begin{equation}\label{eq:PhotonTimeDelay}
    \Delta t = t_2 - t_1 \simeq - \eta_n  \frac{n+1}{2}  \frac{E_\mathrm{2}^{n} - E_\mathrm{1}^{n}}{E_{\mathrm{QG,}n}^{n}}  D_{n}(z_{\mathrm{s}}).
\end{equation}
where $t_1$ and $t_2$ designate the times of flight of photons of energies $E_1$ and $E_2$, respectively. 
Notice that the time delay can be either positive or negative. E.g., for superluminal behaviour, and $E_2 > E_1$, $t'_2$ will be shorter than $t'_1$, and $\Delta t'$ will be negative. $D_{n}(z_{\mathrm{s}})$ accounts for the distance of the source. The expression most commonly used in experimental tests was proposed in~\cite{Jacob:2008bw}:

\begin{equation}\label{eq:LIVComovingDistance}
    D_{n}(z_{\mathrm{s}}) = \frac{1}{H_0} \int\limits_{0}^{z_{\mathrm{s}}}     \frac{(1+z)^{n}}{\sqrt{\Omega_{\mathrm{m}}\left(1+z\right)^{3}+\Omega_{\Lambda}}} dz.
\end{equation}
$H_0$, $\Omega_{\mathrm{m}}$, and $\Omega_{\Lambda}$ are the Hubble constant, the matter density parameter, and the dark-energy density parameter. 
This expression was derived from LIV. DSR can result in different distance contributions, depending on the assumptions under which it was obtained. One of the results in the DSR framework and the LIV contribution, as well as the sensitivity of experimental tests to each of them were compared in~\cite{Bolmont:2022yad}. It is the only study performed so far in which different phenomenological models are compared on experimental data.

DSR with a specific choice of parameters can lead to expression~(\ref{eq:LIVComovingDistance}). Details of dependence on the redshift notwithstanding, $D_{n}(z_{\mathrm{s}})$ serves as a natural amplifier for time delay.

\paragraph{\textbf{Vacuum birefringence}}
Certain aspects of LIV allow photon group velocity to depend not only on the photon energy, but also on the polarisation. This effect is known as vacuum birefringence. In this scenario, the modified photon dispersion relation will be given as
\begin{equation}
    E_{\pm}^2 = p^2c^2 \left(1 \pm  \frac{p_c}{E_{\mathrm{QG,}1}} \right),
\end{equation}
where $\pm$ represent different circular polarisation states. This will lead to different propagation velocities for different polarisation states, which will finally result in a rotation of the polarisation vector of a linearly polarised wave. The rotation angle will depend on the energy of the photon and the distance to the source
\begin{equation}
    \Delta \theta \simeq \frac{E^2}{E_{\mathrm{QG,}1}} D_{1}(z_{\mathrm{s}}),
\end{equation}
where $E$ denotes the measured energy of the photon, and $D_{1}(z_{\mathrm{s}})$ is given in Equation~(\ref{eq:LIVComovingDistance}) for $n=1$.
Photon birefringence arises in the Standard Model extension from mass dimension five operators, which corresponds to $n=1$ modification of the photon dispersion relation. For $n=2$, both polarisation states will have the dispersion relation modified in the same way. 

\paragraph{\textbf{Anomalous gamma-ray absorption}}
Modifying a particle dispersion relation can affect its interactions, either through modifying the kinematics or dynamics of a process, or both. 
This can lead to different reaction thresholds and process rates (i.e. decay widths and cross sections) compared to the ones predicted within the Standard Model. Such processes, most relevant for astrophysics, are synchrotron radiation, Compton scattering, Breit--Wheeler and Bethe--Heitler process, etc. Other phenomena, such as vacuum Cherenkov radiation, photon decay, or photon splitting, are kinematically forbidden under standard special relativity, but become possible if Lorentz symmetry is broken. 

As we already argued in Section~\ref{sec:intro}, not all these effects will be present in DSR. Take photon stability as an example: if there is no preferred frame of reference (as is the case in DSR), photons should decay in all frames or none. Otherwise, one observer would see the process take place, while an observer in another equivalent frame would see it as forbidden. This situation is contradictory to requirement of equivalent inertial frames. 
In addition, DSR is still in its development phase, and the formality is still not fully understood. Therefore, at this stage, one still cannot calculate cross sections. 

Another important aspect to note is that photons are not the only particles involved in these processes. Most commonly they will also involve electrons, which poses the question of whether all particles are equally affected by LIV. 

Finally, while these effects refer to different processes, they all have a similar net result, which is changing a number of emitted or detected gamma rays at different energies. Therefore, measuring these effects boils down to measuring spectra of astrophysical sources and searching for anomalous behaviour.
Let us discuss details with an example. 
The universe is filled with low-energy background electromagnetic radiation, usually classified as radio background (RB), cosmic microwave background (CMB), and extragalactic background light (EBL). 
Gamma rays of very high energies can interact with the photons from the background fields to create electron--positron pairs and are therefore absorbed in these processes. As a rule of thumb in the  special-relativistic scenario, the higher energy of a gamma ray, the more likely it is to be absorbed, which results in the softening of the spectrum. The optical depth of the universe to the gamma rays is given as:

\begin{equation}\label{eq:optical_depth}
\tau(E,z_{\mathrm{s}})=\int_{0}^{z_{\mathrm{s}}}\frac{dl}{dz} d z \int_{-1}^{1} \frac{1-\cos\theta^{\prime}}{2}  d\cos\theta^{\prime}\int_{\epsilon_{th}^{\prime}}^{\infty} \sigma_{\gamma \gamma}\left(s\right) n\left(\epsilon^{\prime}, z\right) d \epsilon^{\prime}.
\end{equation}
It depends on the comoving number density of background photons per unit energy, here marked as $n\left(\epsilon^{\prime}, z\right)$, and the cross section given in standard quantum electrodynamics as
\begin{equation}\label{eq:BWcrossSection}
    \sigma_{\gamma \gamma}(s) = \frac{2\pi\alpha^2}{3m_{\mathrm{e}}^2} \left( 1-\beta^2 \right) \left[ 2\beta \left(\beta^2 - 2\right) + \left(3 - \beta^4\right) \ln{\left(\frac{1+\beta}{1-\beta}\right)} \right],
\end{equation}
with 
\begin{equation}\label{eq:beta}
    \beta(s) = \left[ - \frac{4m_{\mathrm{e}}^2 c^4}{s} \right]^{1/2},
\end{equation}
and $s$ being the invariant mass. The integrals run (right to left) over all background photon energies, $\epsilon^{\prime}$, scattering angles, $\theta^{\prime}$, and the thickness of the medium between the source and the observer. Primed variables are given in the interaction comoving frame. Therefore, $\epsilon_{\mathrm{th}}^{\prime}$ denotes reaction threshold for the Breit--Wheeler process in the comoving frame.
The expression for $dl/dz$ depends on the adopted cosmological model. For the $\Lambda$CDM cosmology, it is given as:
\begin{equation}\label{eq:ComovingDistance}
\frac{dl}{dz} = \frac{c}{H_{0}\left(1+z\right) \sqrt{\Omega_{\mathrm{m}}\left(1+z\right)^{3}+\Omega_{\Lambda}}}.
\end{equation}

\begin{figure}
\centering
\includegraphics[width=0.48\textwidth]{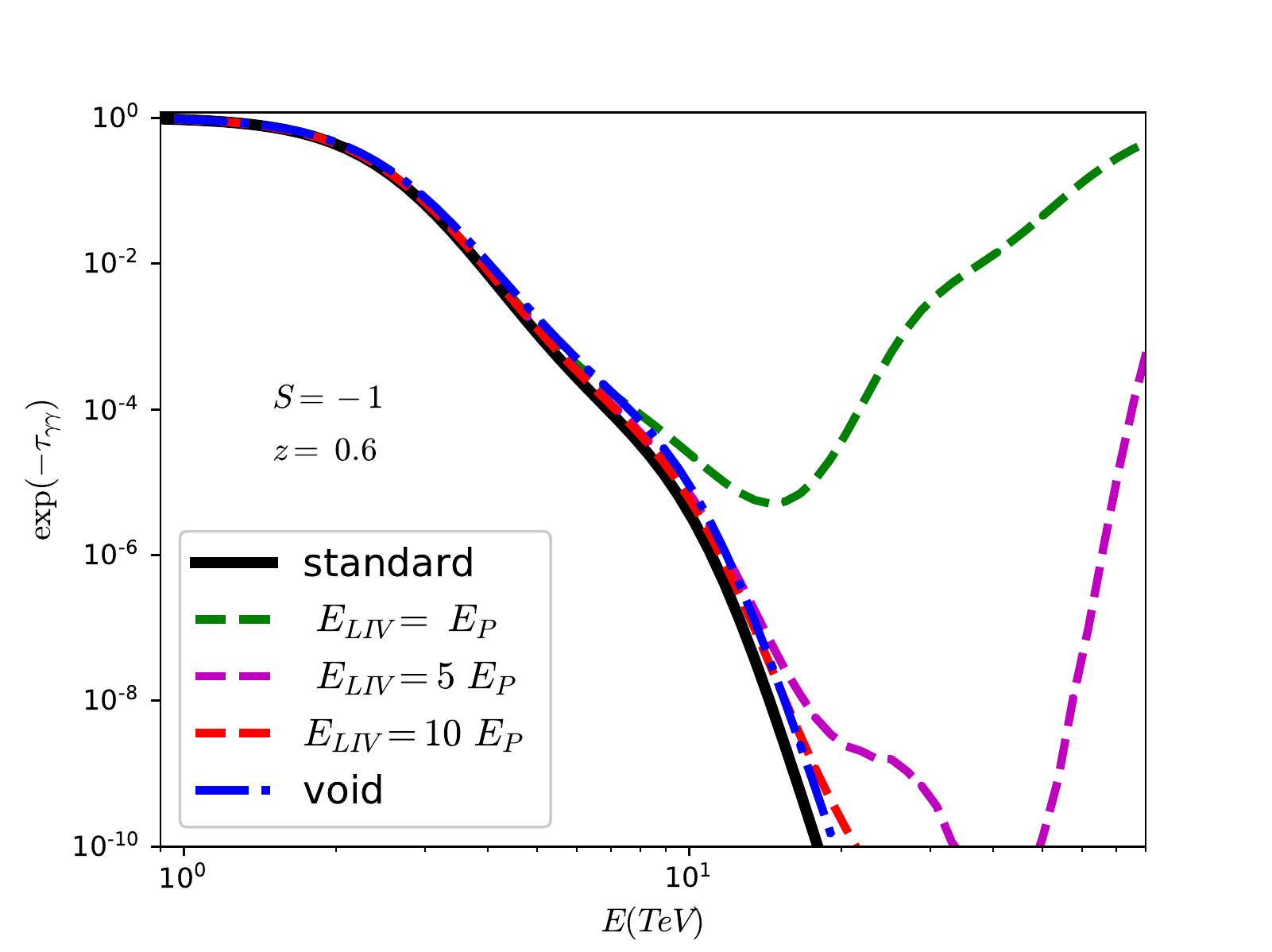}
\includegraphics[width=0.48\textwidth]{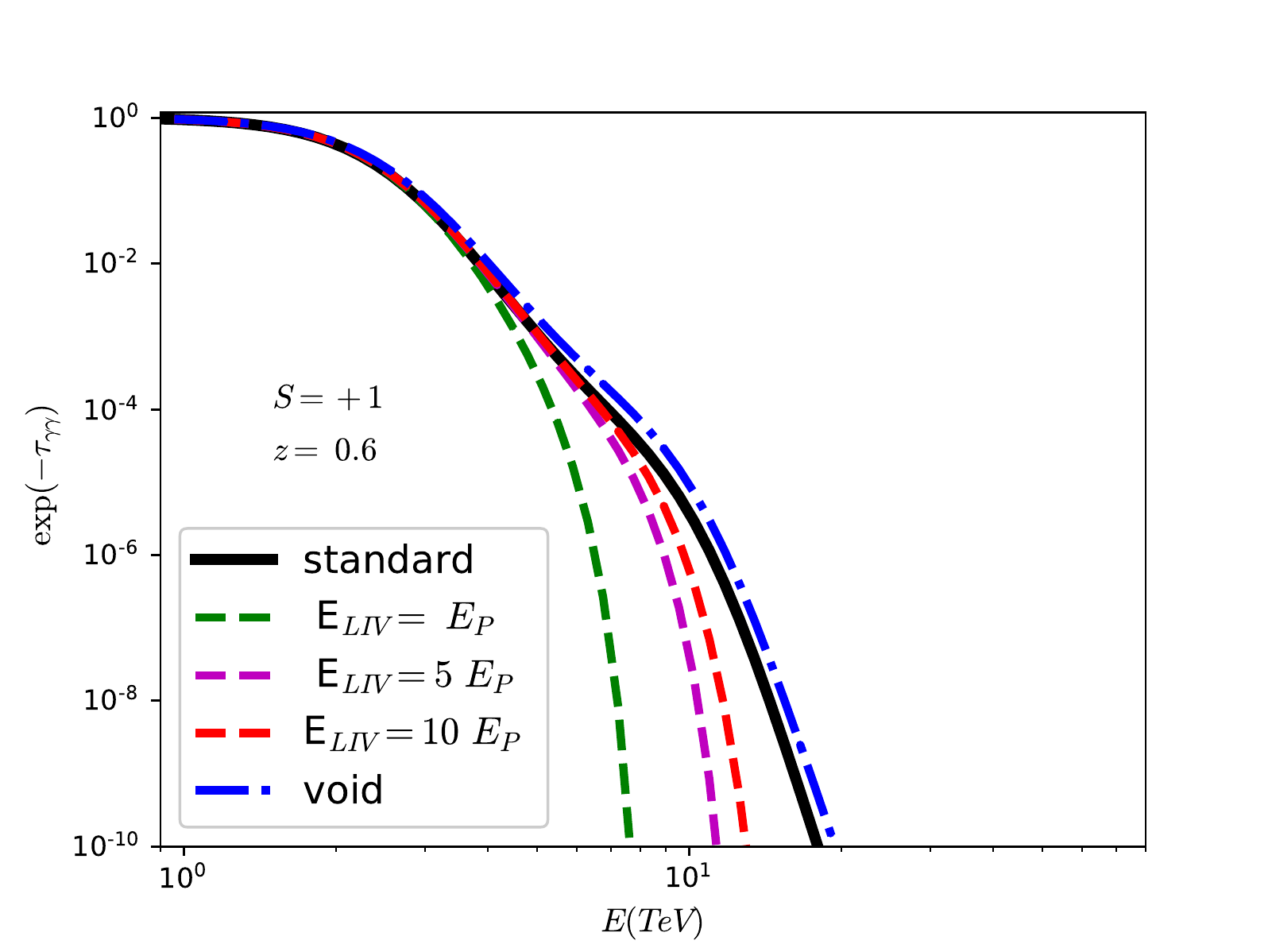}
\caption{Gamma-ray absorption coefficients as a function of gamma-ray energy for a source at redshift $z_\mathrm{s} = 0.6$. The subluminal (here marked by $S=-1$) and superluminal (here marked by $S=+1$) scenarios are represented in the left and right plots, respectively. The standard special-relativistic scenario is represented by solid black lines in bot plots, while dashed lines represent modified absorption for different values of quantum gravity energy scale. In all cases $n=1$ modification was considered. The dot-dashed blue line represents another effect investigated by the authors not connected to LIV. Figure adopted from \cite{Abdalla:2018sxi}. Reproduced by permission of the AAS.}
\label{fig:EBLattenuation}
\end{figure}

Violation of Lorentz symmetry can result in the modification of the kinematics or dynamics of the process. It can be relatively easy to show that the reaction threshold changes to~\cite{Blanch:2001hu}
\begin{equation}\label{eq:modified_energy_threshold}
\epsilon_{\mathrm{th}}^{\prime}= \frac{2 m_{\mathrm{e}}^{2} c^{4}}{E^{\prime} (1-\cos\theta^{\prime})} - \frac{\eta_n}{2 (1-\cos\theta^{\prime})}\left(\frac{E^{\prime}}{E_{\mathrm{QG,}n}}\right)^{n}E^{\prime},
\end{equation}
while the invariant mass becomes 
\begin{equation}\label{eq:modified_center_of_mass}
s= 2 E^{\prime} \epsilon^{\prime} (1-\cos\theta^{\prime})+ \eta_n \left(\frac{E^{\prime}}{E_{\mathrm{QG,}n}}\right)^{n} E^{\prime 2}.
\end{equation}
In both cases, the first term is the same as in the special relativity, while the second term is a consequence of the modification of the photon dispersion relation. 
The net result of these modifications is that, depending on the value of $\eta_n$, the gamma-ray absorption by the background fields will be weaker (for $\eta_n = -1$, i.e. subluminal behaviour), or stronger (for $\eta_n = +1$, i.e. superluminal behaviour) than in standard special-relativistic case. In other words, the universe will be most transparent to gamma rays in subluminal scenario, and least transparent in the superluminal scenario. Modified gamma-ray absorption is shown in Figure~\ref{fig:EBLattenuation}. 
This phenomenon of modified universe transparency will be reflected on the observed spectra of astrophysical sources. To be specific, superluminal scenario will result in the observed spectra to appear even softer than the standard absorption. There is no clear way of resolving the effect of LIV in the superluminal scenario from what is expected in standard physics. Subluminal behaviour, on the other hand, will lead to harder spectra. On top of that, at high enough gamma-ray energies, the absorption will cease entirely, and provided there is emission at those energies, the spectrum will correspond to the source intrinsic one. These features could be detected in the spectra, provided the source emits gamma rays at high enough energies.

We will not discuss here other possible consequences of LIV in so many details. However, we will say that, if real, they have similar effects on the observed spectra. For example, in the superluminal scenario the photon can become unstable and decay into an electron--positron pair, or split in several lower-energy photons, where splitting in three photons is the most dominant channel. As a result, not all emitted gamma rays will reach the detector. The effect is similar to the gamma-ray absorption by the EBL, especially in the superluminal case. However, photon decay and splitting would be seen as a cutoff in the spectra, rather than a gradual attenuation. 
Just as Breit--Wheeler process can be modified in LIV, so can the Bethe--Heitler, which describes gamma-ray interactions with an atomic nucleus. To be specific, the cross section in the subluminal scenario is significantly smaller, while in the superluminal scenario it remains quite similar to the standard special-relativistic one. In either case the number of gamma rays reaching Earth will not change with respect to standard special relativity. However, a smaller cross section means that particle showers will develop deeper in the atmosphere. Consequently, fewer showers will be detected, resulting in the observed spectra resembling the scenario of superluminal EBL absorption.

\subsubsection{Analysis methods}
\label{sec:Gamma-rayAnalysisMethods}

Several analysis methods have been proposed to test energy-dependent photon group velocity. Here we will focus on the maximum likelihood (ML) estimation method\footnote{An interested reader can find description and comparison between different methods in \cite{Terzic:2021rlx} and references therein.}, recently most often employed by experimentalists. The maximum likelihood estimation is a powerful statistical method applicable to various problems. 

We will first discuss the application in the time of flight measurements, for which this method was developed and first introduced in~\cite{Martinez:2008ki}. 
We start by defining a probability density function (PDF) for a photon of energy $E$ to be detected at a time $t$, under certain conditions:

\begin{equation}\label{eq:PDF}
    f^{(\mathrm{s})}(E, t) = \int_0^{\infty} G \left( E, E_{\mathrm{true}} \right) \, A_{\mathrm{eff}}(E_{\mathrm{true}}, t) \, \Phi(t', E_{\mathrm{true}}) \, dE_{\mathrm{true}}.
\end{equation}
All our knowledge and assumptions about the emission process  and the measurement technique are contained in this expression. In particular: \\
\textbf{Energy resolution and bias}, $G \left( E, E_{\mathrm{true}} \right)$, takes into account imperfections of the instrument. It is the probability for a gamma ray of energy $ E_{\mathrm{true}}$ to be reconstructed as $E$. For this reason, the integral goes over all possible values of $ E_{\mathrm{true}}$. \\
\textbf{Instrument acceptance}, $A_{\mathrm{eff}}(E_{\mathrm{true}}, t)$, sometimes also called \textit{effective area} or \textit{collection area}, is the probability for the instrument to detect a gamma ray of energy $E_{\mathrm{true}}$ at a moment $t$. This function will be null for energies outside of the instrument sensitivity range. Dependence on time represents changing observation conditions.\\
\textbf{Emission and propagation effects} are contained in $\Phi(E_{\mathrm{true}}, t')$. This includes the energy and temporal distribution of gamma rays at the emission, as well as propagation effects. These include redshift of the gamma-ray energy, gamma-ray absorption on the background radiation, or effects induced by LIV. 
The emission time $t'$ is related to the detection time $t$ in the following way:
\begin{equation}\label{eq:t_prime}
    t' = t + \xi_{n}E_{\mathrm{true}}^{n},
\end{equation}
where the parameter $\xi_{n}$ represents the strength of the LIV effect. It is introduced in the following way to facilitate numerical computations:
\begin{equation}\label{eq:eta_n}
    \xi_n = - \eta_n  \frac{n+1}{2}  \frac{1}{E_{\mathrm{QG,}n}^{n}}  D_{n}(z_{\mathrm{s}}).
\end{equation}

Obviously, the probability of detecting a photon of a certain energy at a given time will depend on the assumed distribution of gamma rays at the emission, the so-called emission template. While gamma-ray detectors are capable of precise measurements of arrival times for each photon, our ability to estimate the emission time for each photon emitted from an astronomical object is limited at best. Therefore, in order to make precise measurements on LIV, the emission time needs to be constrained in some way. Unlike in laboratory experiments, astrophysicists cannot control their sources. But we can choose the most adequate ones for a particular study. In this case, sources with strong and fast changes of flux make a good choice. Sources of highly variable gamma-ray flux are pulsars, gamma-ray bursts (GRB), and active galactic nuclei (AGN) in high-emission states, usually called flares.\\ 
Pulsars, with their millisecond pulses, provide a very strong constraint on the emission time, and are very reliable. On the other hand, only a handful of pulsars have been detected at energies of a few 100\,GeV and above, and all of them inside the Milky Way, thus the LIV effects, if any, are less amplified by the distance. Gamma-ray bursts are violent explosions, usually associated with collapses of massive stars into black holes or mergers of binary neutron stars. These transient events are energetic enough to be seen in other galaxies and at large redshifts, and their short duration and variability provide strong constraints on the emission time. However they are entirely unpredictable. Instruments onboard the \textit{Fermi} satellite are capable of detecting numerous GRBs because of their wide field of view. Cherenkov telescopes, on the other hand, observe only small portions of the sky at the time, and rely on other instruments to alert them of ongoing GRBs. However, repointing takes time, making GRBs notoriously difficult to catch. In addition, the signal above $\sim100$\,GeV from GRBs at redshifts above $z_\mathrm{s} = 1$ is attenuated because of gamma-ray absorption on extragalactic background light, posing an additional difficulty. So far, only four GRBs were significantly detected with Cherenkov telescopes, and only one was used for LIV study. We will take a closer look into that case shortly.
Active galactic nuclei are persistent strong sources of gamma rays at distances from very small to large redshifts. However, steady emission is not particularly useful for time of flight studies. Flaring episodes, on the other hand, are characterised by fast changes of flux, which constrain the emission time and also provide richer data samples of gamma rays. Although the flux variability in AGN flares is not as fast as in pulsars or GRBs, these states of enhanced emission last longer than GRBs, can sometimes be predicted from observations in lower-energy bands, and are in general easier to catch. 

\begin{figure}
\centering
\includegraphics[width=0.65\textwidth]{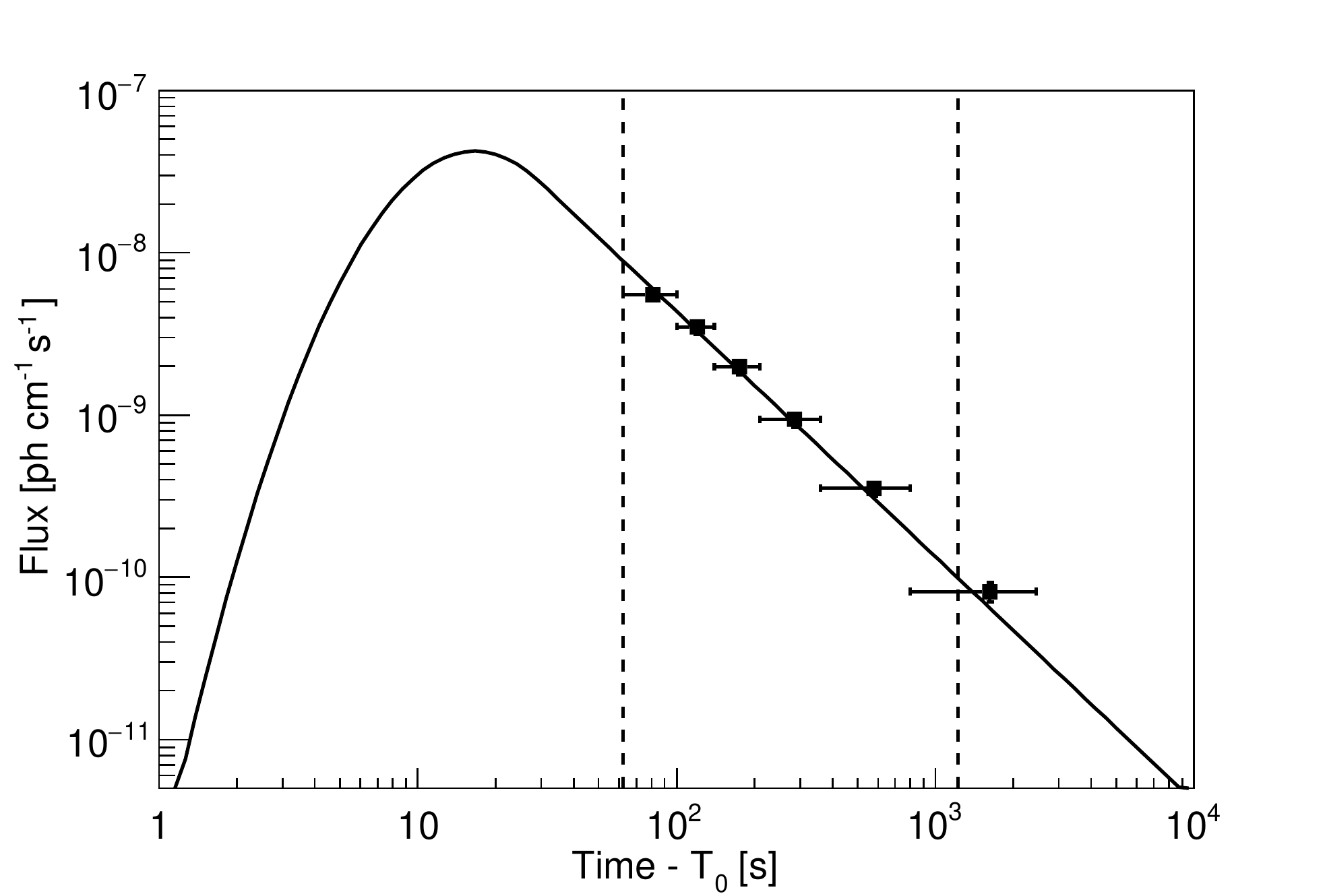}
\caption{Light curve of GRB\,190114C above 300\,GeV. Black points represent measured flux by the MAGIC telescopes. The full black line represents the emission template as reported in \cite{MAGIC:2019irs}. Reprinted figure with permission from from \cite{2020PhRvL.125b1301A}. Copyright 2020 by the American Physical Society.}
\label{fig:GRB190114C_LCtemplate}
\end{figure}

Let us consider an example of LIV study performed on gamma-ray burst GRB\,190114C observed by the MAGIC telescopes \cite{2020PhRvL.125b1301A}, which was the first such study done on a gamma-ray data observed by Cherenkov telescopes.
The MAGIC telescopes detected GRB\,190114C above 1\,TeV in energy~\cite{MAGIC:2019lau}. The observations started 62 seconds after the burst, with the light curve showing smooth decay of the flux. The MAGIC observations results are shown in Figure~\ref{fig:GRB190114C_LCtemplate} with black points. 
Note that the scales are logarithmic, meaning that the flux changes very quickly. However, the decay is a monotone power law. One can easily demonstrate that adding an energy-dependent time delay to individual gamma rays will not change the overall shape of the light curve, therefore making it virtually impossible to resolve source-intrinsic effects from the ones induced by LIV. Fortunately, experts in gamma-ray bursts managed to create model of the gamma-ray emission based on observations in lower energy bands combined with theoretical inferences \cite{MAGIC:2019irs}. The result is represented with a full black line in Figure~\ref{fig:GRB190114C_LCtemplate}. Using this model as a template for the temporal distribution of events provides a very strong handle on energy-dependent time delay. A very sharp peak in flux will strongly dominate the probability density function (Equation~\ref{eq:PDF}). In this case, no change of spectrum with time was detected, so the entire emission template was obtained by folding the light curve with the observed spectrum. 

Once we have the probability density function for each event in the data sample, we combine them in the likelihood function
\begin{equation}\label{eq:Likelihood}
    \mathcal{L}(\xi_{n})= \prod_{i=1}^{N_{\mathrm{ON}}} \left(p^{(\mathrm{s})}_{i}  \frac{f^{(\mathrm{s})}(E_{i}, t_{i})}{\int_{E_{\mathrm{min}}}^{E_{\mathrm{max}}} dE \int_{t_{\mathrm{min}}}^{t_{\mathrm{max}}} f^{(\mathrm{s})}(E, t) dt} + p^{(\mathrm{b})}_{i} \frac{f^{(\mathrm{b})}(E_{i}, t_{i})}{\int_{E_{\mathrm{min}}}^{E_{\mathrm{max}}} dE \int_{t_{\mathrm{min}}}^{t_{\mathrm{max}}} f^{(\mathrm{b})}(E, t) dt}\right).
\end{equation}
The product goes over all events in the signal region. As noted in Section~\ref{sec:Gamma-rayDetectors}, the signal region contains also some background events, which also need to be taken into account. The superscripts "s" and "b" stand for signal and background, respectively. $p^{(\mathrm{s})}_{i}$ is the probability that event $i$ belongs to the signal, and respectively for background. The probability density function for background ($f^{(\mathrm{b})}(E_{i}, t_{i})$) is similar to the one for signal, given in Equation~(\ref{eq:PDF}). The energy and temporal distributions of the background events is not necessarily the same as for the signal, while the instrument effects should be the same. However, the most important difference is that we do not consider a time delay in background. While it may be present, we do not exactly know the sources of the background effects, so we cannot determine the event distributions at the source; we only consider them at the detector.
Putting it simply, we do not care what happens with the background, we just acknowledge that it is present.

Now that we have all the components of the likelihood function, we maximise it for $\xi_n$. To be more precise, we define test statistics (TS) as
\begin{equation}
    L(\xi_n) = -2 \ln{\frac{\mathcal{L}_0(\xi_{n})}{\mathcal{L}(\xi_{n})}},
\end{equation}
where $\mathcal{L}_0$ corresponds to the likelihood function for the null model. According to the Wilks' theorem, the test statistics asymptotically approaches the $\chi^2$ distribution under the null hypothesis \cite{10.1214/aoms/1177732360}. 
The value of $\xi_n$ for which $L$ has minimum is the estimation of the true value that we are looking for. This procedure enables us also to determine the statistical significance of the measured value of $\xi_n$, as well as to produce confidence intervals. 
The strongest constraints on the energy-dependent photon group velocity for the order $n=1$ were set based on the observation of GRB\,090510 with \textit{Fermi}-LAT to $E^{(-)}_\mathrm{QG,1} > 2.2\times 10^{19}$\,GeV and $E^{(+)}_\mathrm{QG,1} > 3.9\times 10^{19}$\,GeV, for subluminal and superluminal scenario, respectively~\cite{Vasileiou:2013vra}. 
As discussed in \cite{Terzic:2021rlx}, the sensitivity to quantum gravity energy scales is some what different for orders $n=1$ and $n=2$ (see Table~\ref{Tab:TOFsensitivity}).

\begin{table}[!t]
\caption{\label{Tab:TOFsensitivity}Sensitivity to $E_{\mathrm{QG,}n}$ considering characteristics of the source and the sample. $E_{\mathrm{max}}$ is the highest gamma-ray energy in the sample, $t_{\mathrm{var}}$ is the shortest variability timescale in the light curve, and $z_{\mathrm{s}}$ is the redshift of the source}
\begin{tabular}{p{2cm}p{2.4cm}p{2cm}p{2.4cm}p{2.cm}}
\noalign{\smallskip}
\noalign{\smallskip}\hline\noalign{\smallskip}
$E_{\mathrm{QG,1}}$ &  $\propto$ &$E_{\mathrm{max}}$ &$t_{\mathrm{var}}^{-1}$ &$z_{\mathrm{s}}^{\sim1}$ \\
\noalign{\smallskip}
$E_{\mathrm{QG,2}}$ &  $\propto$ &$E_{\mathrm{max}}$ &$t_{\mathrm{var}}^{-1/2}$ &$z_{\mathrm{s}}^{\sim 2/3}$\\
\noalign{\smallskip}\hline\noalign{\smallskip}
\end{tabular}
\end{table}

\noindent
Thanks to the very fast change of flux and relatively large redshift of GRB\,090510 the sensitivity to the 1st order modification was very high. When it comes to the 2nd order modification, the maximal energy in the sample carries more weight compared to other two parameters. That is where Cherenkov telescopes have the advantage over space detectors, and why the strongest constraints for the order $n=2$ were set based on the observation of the blazar Mrk\,501 with the H.E.S.S. telescopes: $E^{(-)}_\mathrm{QG,2} > 8.5\times 10^{10}$\,GeV and $E^{(+)}_\mathrm{QG,2} > 7.3\times 10^{10}$\,GeV, for subluminal and superluminal scenario, respectively~\cite{Abdalla_2019}. 
The GRB\,190114C had all three requirements met to a high degree, and while the LIV constraints were not the most constraining in either scenario, they were very close and supporting the results from the studies performed on GRB\,090510 and Mrk\,501.

From the experimental point of view, the maximum likelihood method is a very natural tool because it allows to directly include the information on the detector response in the analysis. Moreover, any other unknown can be introduced as a nuisance parameter. 
On the other hand, one obvious downside of this approach is that it requires making certain assumptions about the emission processes and propagation of gamma rays. 
Our present knowledge of astrophysical sources is not nearly good enough to predict the exact emission time of each particular photon. Moreover, it is quite possible that the source-intrinsic processes are energy correlated, which could mimic or disguise effects of LIV. 
It will still be quite some time before our understanding of sources emission mechanisms becomes precise enough to describe correlations. For the time being, a clever workaround is to combine sources at different distances. While the time delay coming from LIV clearly depends on the distance between the source and the detector, the source intrinsic processes are not expected to be distance dependent. Furthermore, different types of sources have different emission mechanisms, so by combining them in a single analysis the effect of possible correlations in source-intrinsic processes is decreased. This method was explored in~\cite{Bolmont:2022yad}. So far, only a proof of concept was demonstrated on Monte Carlo simulations, with a study on real data promised to follow soon.

Probability density, likelihood, and test statistics, as statistical methods, are also used to search for other effects of LIV.
Many astrophysical sources emit polarised photons. Provided that the polarisation of emitted light is known, one can measure the birefringence effect by comparing polarisation of the detected photons to the expected polarisation at the emission. However, even if there is no way of establishing the angle of polarisation at the emission, some processes, such as synchrotron radiation, emit strongly polarised photons. If the angle of rotation of the polarisation depends on the photon energy, after crossing astrophysical distances, the signal will be depolarised. Therefore, one can measure the degree of polarisation in the signal against expected degree of polarisation when no LIV is assumed. Such study was performed in \cite{PhysRevD.102.043008}, using broadband optical polarimetry of 1278 AGN and GRBs.  
Unfortunately, it is virtually impossible to measure gamma-ray polarisation with detection techniques which rely on particle showers. The highest energies used to measure vacuum birefringence were a few hundred keV. As modest as this may be compared to the PeV energies accessible to us nowadays, depolarisation measurements placed very strong constraints on the energy-dependent polarisation hypothesis.

As explained in Section~\ref{sec:LIVeffects}, subluminal scenario can induce measurable features in the observed spectra of gamma-ray sources. Such features have not been detected so far, but strong constrains have been set to the quantum gravity energy scales. 
As in the time of flight studies, we do not know exactly the intrinsic spectra of sources that we observe. However, we do have certain expectations based on our knowledge of our sources. 
\begin{figure}
\centering
\includegraphics[width=0.65\textwidth]{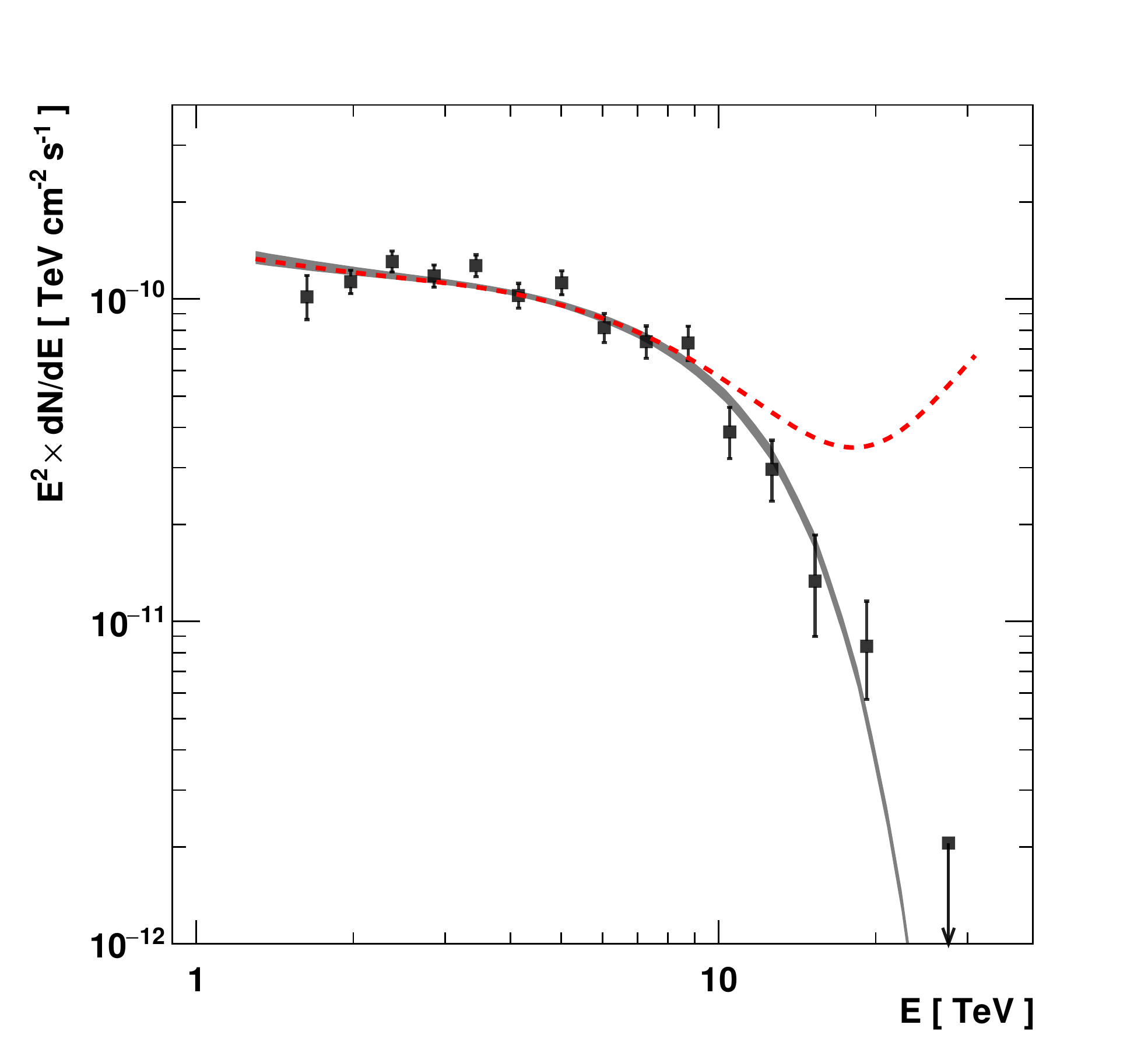}
\caption{Spectral energy distribution of blazar Mrk\,501 observed with the H.E.S.S. telescopes. The black points and full black line represent the fit to the EBL-attenuated spectrum. A simple power law was assumed for the intrinsic spectrum. The expected spectrum observed for the same intrinsic shape but considering subluminal scenario with $n=1$ and the quantum gravity energy scale corresponding to the Planck energy is represented by the dashed red line. Figure adopted from \cite{Abdalla_2019}. Reproduced by permission of the AAS.}
\label{fig:Abdalla_2019_spectrum}
\end{figure}
In universe transparency studies, as in the case of active galactic nucleus Mrk\,501 observed with the H.E.S.S. telescopes shown in Figure~\ref{fig:Abdalla_2019_spectrum}, one starts by assuming a source-intrinsic spectrum. In this case, a simple power law was assumed. Then, an absorbed spectrum is calculated using a certain EBL model (in this case, the authors decided on the one from \cite{2008A&A...487..837F}) and assuming no LIV, and fitted to the data. The resulting spectrum is shown with a black line in Figure~\ref{fig:Abdalla_2019_spectrum}. Then the procedure is repeated for different values of $E_\mathrm{QG}$ to test hypotheses of different levels of LIV. For each value of $E_\mathrm{QG}$ test statistics is computed according to
\begin{equation}
    TS = \chi^2(E_\mathrm{QG}) - \chi^2(E_\mathrm{QG} \rightarrow \infty),
\end{equation}
where $E_\mathrm{QG} \rightarrow \infty$ corresponds to no LIV. Putting all test statistics together creates a test statistics profile, which, in a similar manner as in the time of flight studies, allows estimating confidence intervals and constraining the quantum gravity energy scales. 

In the study described here, it was assumed that electron dispersion relation is not modified, and the only effect of LIV is modified kinematics of the Breit--Wheeler process. How solid are these assumptions? If we had to take into account modifications of dispersion relations of several particles, and consider different LIV effects on top of that, it would be virtually impossible to obtain a very significant result. Fortunately, there are other types of studies which constrain these other effects, allowing us to neglect them. Let us start with the modification of electron dispersion relation. The modifying term in the electron dispersion relation for $n=1$ was constrained based on the synchrotron radiation from the Crab nebula to seven orders of magnitude above the Planck energy~\cite{Jacobson:2002ye}.
Compton scattering is one of the most important processes for production of gamma rays in AGN. LIV could also result in anomalous Compton scattering, which would compete with other processes and effects, making them extremely difficult to resolve. However, based on several combined arguments, it was shown that possible LIV effects on the Compton scattering were unlikely to be relevant in realistic astrophysical environments~\cite{Abdalla:2018sxi}. 
Photon decay and splitting is only possible in the superluminal scenario, with no equivalent process in the  subluminal scenario. As we argued earlier, gamma-ray absorption in the superluminal scenario becomes even stronger, and very difficult to resolve from source-intrinsic effects, which is why anomalous absorption is usually tested in the subluminal scenario. In that case, there is no danger of photon instability contaminating the results. However, if electromagnetic shower development in the atmosphere was modified in subluminal LIV scenario, it would have the opposite effect to the subluminal LIV gamma-ray absorption, and the two effects would compete against each other. It was, therefore, essential to test for one of these effects in an environment where the other was not present. The effect on the shower development was tested on the gamma rays from the Crab Nebula~\cite{Rubtsov:2016bea}. The Crab Nebula is located $2.0 \pm 0.5$\,kpc from Earth, which is a too short distance for gamma-ray absorption on EBL to be significant, providing a clean data sample to constrain anomalous shower development. Influence of LIV on the extensive air shower development was constrained to $E_\mathrm{QG,2} > 2.1\times 10^{11}$\,GeV (note that only order $n=2$ was constrained).
Considering all these constraints, gives us ground to test anomalous gamma-ray absorption independently. 
As in the time of flight studies, multiple sources can also be used to test the anomalous gamma-ray absorption. Eighteen spectra from six different AGN were used in \cite{Lang:2018yog} to set the strongest constraints on universe transparency to date: $E_\mathrm{QG,1} > 6.9\times 10^{19}$\,GeV and $E_\mathrm{QG,2} > 1.6\times 10^{12}$\,GeV.

\subsection{Neutrinos}
\label{subsec:NUs}
Neutrinos are excellent cosmic messengers since they are electrically neutral and interact only weakly (and gravitationally). They can therefore reach us without deflection or appreciable absorption directly from their sources. But they are very difficult to detect and that poses an experimental challenge. The ingenuity of experimentalists has made it possible to build large neutrino telescopes that can collect enough events to extract statistically meaningful measurements. 
Neutrino telescopes are large-volume detectors using open transparent media like water or ice both as a target and as Cherenkov medium.  They detect the Cherenkov radiation emitted by particles produced in high-energy neutrino interactions and are able to reconstruct the original direction and energy of the neutrino from this information~\cite{Anton:2020veo}. They are deployed at great depths to reduce the copious flux of muons produced in cosmic ray interactions in the atmosphere that leave tracks in the detectors. Although the main reason to build  neutrino telescopes is to study the universe with high-energy neutrinos, were high-energy refers to neutrino energies above a few tens of GeV and up to several PeV, they have proved to be highly versatile detectors capable of addressing many topics related to fundamental physics in a competitive way to accelerator experiments~\cite{Ahlers:2018mkf,Margiotta:2019agz,Ackermann:2019cxh,PerezdelosHeros:2020tan}.

The detector sizes, typically $\cal{O}$(km$^3)$, are determined by the weak astrophysical neutrino flux they are built to study and by the tiny neutrino cross section with matter. 
There are currently three large neutrino telescopes in operation, IceCube~\cite{IceCube:01, IceCube:2006tjp} at the geographic South Pole, KM3NET~\cite{KM3Net:01, KM3Net:2016zxf} in the Mediterranean sea and Baikal in lake Baikal~\cite{Baikal:01, Avrorin:2011zza}. An additional project, P-ONE~\cite{P-ONE:01, P-ONE:2020ljt} is under R\&D in the Pacific ocean, off the coast of Canada, while ANTARES~\cite{ANTARES:01, ANTARES:2011hfw}, the predecessor of KM3NET off the coast of Toulon, in the Mediterranean, has recently been decommissioned after sixteen years of operation. 
Smaller underground neutrino detectors with an energy threshold of a few MeV, like Super-Kamiokande~\cite{Super-Kamiokande:01, Super-Kamiokande:2002weg}, are complementary to their high-energy brothers in energy reach and can also competitively address fundamental physics topics.

\begin{figure}[t]
  \begin{minipage}[c]{0.47\textwidth}
    \includegraphics[width=\textwidth]{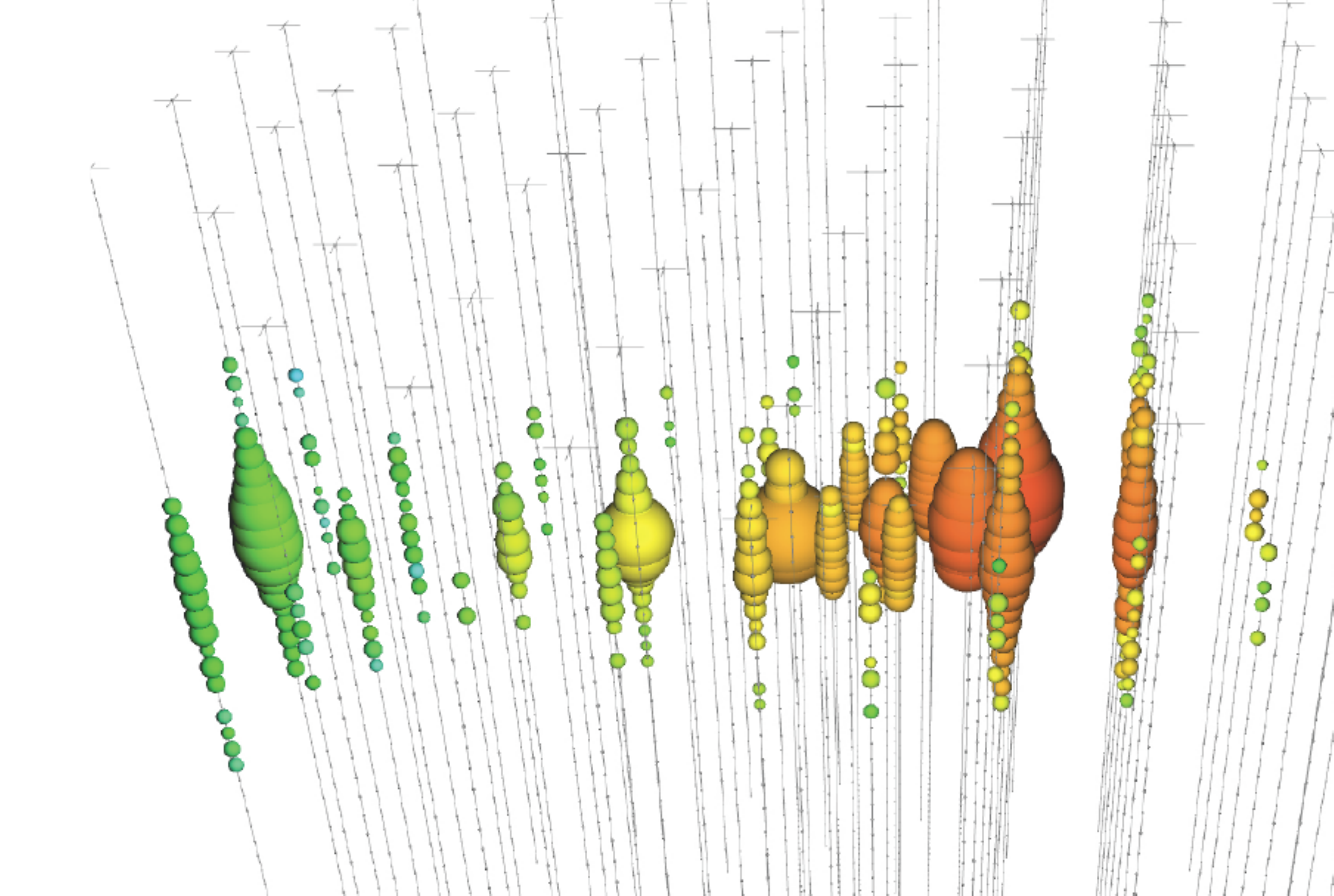}
  \end{minipage}\hfill
  \begin{minipage}[c]{0.47\textwidth}
    \includegraphics[width=\textwidth]{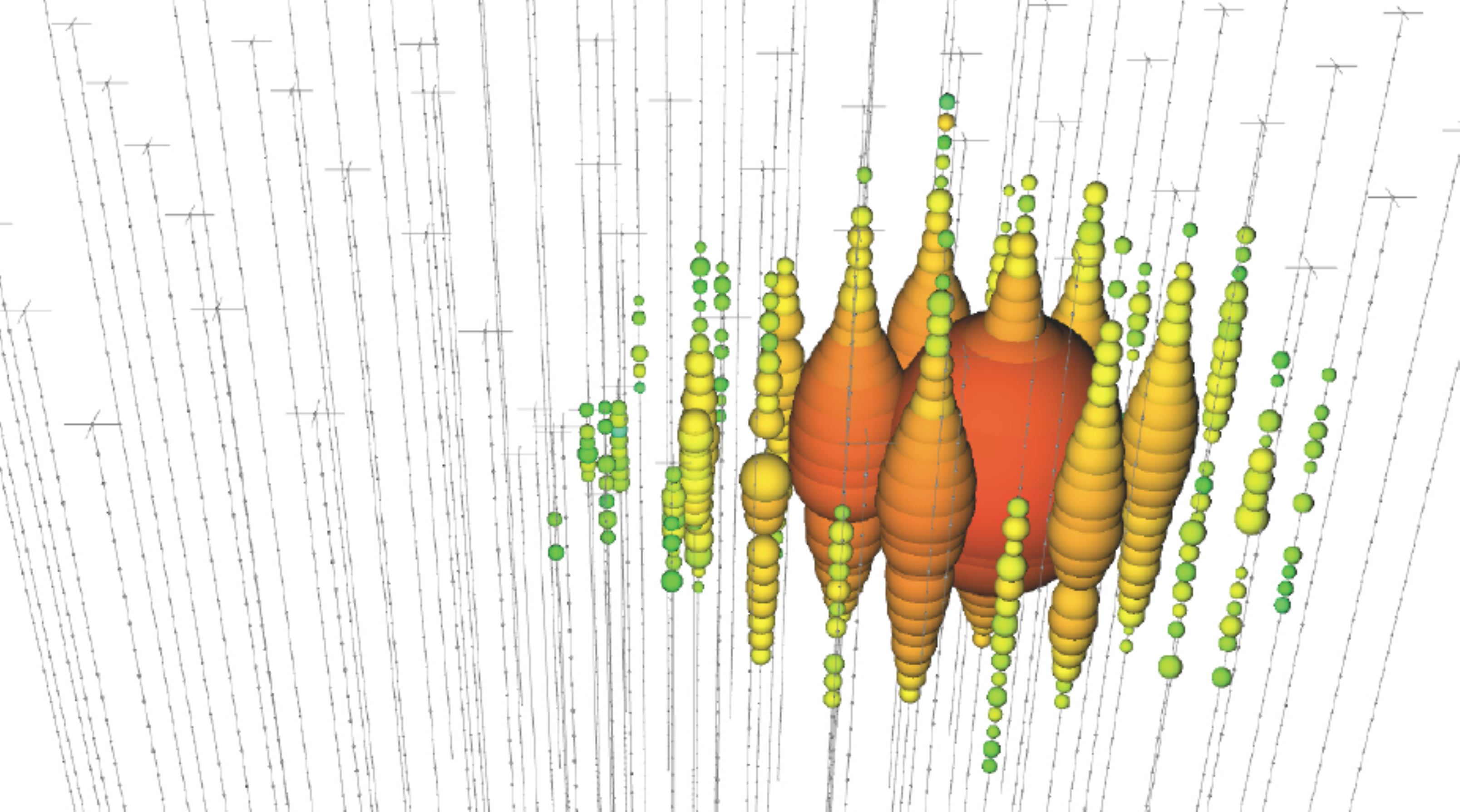}
  \end{minipage}
  \caption{Typical event patterns in a large-scale neutrino telescope. {\bf Left:} A muon track from a $\nu_\mu$ interaction. {\bf Right:} The signature of the cascade of particles produced in a $\nu_e$ or $\nu_\tau$ charged-current interaction or any flavour neutral-current. Each coloured dot represents a hit optical module. The size of the dot is proportional to the amount of light detected and the colour code is related to the relative timing of light detection: red denotes earlier hits, blue corresponds to later hits. Figure from~\cite{Ahlers:2018mkf}.}
    \label{fig:evt_signatures}
\end{figure}

The common design feature of all these detectors is an array of optical modules monitoring a volume of a transparent medium. The exact design of the optical modules depends on the detector, but in essence an optical module consists of one, or several, photomultiplier tubes with associated electronics to timestamp and digitize the detected signal. Charged-current neutrino interactions of any flavour on a nucleon $N$, $\nu_l + N \rightarrow X + l$, will produce the corresponding lepton, $l$, along with a cascade of particles $X$ at the interaction vertex from the hadronization and decays of the interaction products, see Figure~\ref{fig:evt_signatures}. The muon produced in charged-current $\nu_{\mu}$ interactions can travel several kilometers in matter, depending on energy, and will produce a track in the detector which follows the original direction of the neutrino. The electron and tau from $\nu_e$ and $\nu_\tau$ interactions will rapidly lose energy or decay, respectively, not leaving any visible track, only the particle cascade at the vertex. Neutral-current interactions, $\nu_l + N \rightarrow X + \nu_{l}$, where the outgoing neutrino is not detected, will also leave a cascade-type signature in the detector at the interaction point. The absolute time resolution of the optical modules must be of the order of a ns to allow precise event reconstruction, and the detector is calibrated so that the amount of light detected can be translated to the total energy of the neutrino. Additionally, the global timing of the detector is kept synchronized to a master GPS also to the ns level to be able to do astronomy. These measurements, neutrino arrival time, pointing accuracy and energy resolution, define the performance of a neutrino telescope. Typically the energy reconstruction for cascades is better than for events producing a muon (if this one is not contained, i.e., is produced and decays, in the detector volume). The typical energy resolution for cascade events in a neutrino telescope, depending on design, is 10\% or better, while for through-going tracks it can be more than double that. Note that the geometric instrumented volume of a neutrino telescope does not coincide with its effective detection volume. The latter depends on the signal searched for and the event energy, so the effective volume of a neutrino telescope is analysis dependent and it can be both larger or smaller than the geometric volume.

There are several ways to use high energy neutrinos to search for LIV effects and there are excellent reviews on the subject~\cite{Roberts:2021vsi,Torri:2020dec,Stecker:2017gdy,Diaz:2014yva,Stecker:2014oxa,Hooper:2005jp} but we will concentrate here on two techniques that are especially relevant for the topic of this book: measuring the relative arrival times of the different messengers in a multi-messenger observation of a transient event, like a supernova, GRB, or flaring AGN, or to use neutrino oscillations, an interference phenomenon, which can provide a very sensitive probe of new physics.

\paragraph{\textit{Multi-messenger observations}}
In this approach the LIV signature is a deviation in the speed of propagation of energetic neutrinos with respect to electromagnetic radiation in the form of relative time of arrival. The time difference with respect to the standard (no LIV) scenario can be both positive (delay) or negative (pre-arrival) of high energy neutrinos with respect to photons. 
As for studies performed on gamma-rays alone, GRBs and flaring AGN are excellent candidates for these multi-messenger studies as well, see Section~\ref{sec:LIVeffects}.
The effect on the propagation of high energy neutrinos can be parameterized quite generically as a modification of the dispersion relation, as was done in Equation~(\ref{eq:generalMDR}). 
Note that at the energies of relevance, $\cal{O}$(TeV) and above, neutrinos can be considered as massless particles in these studies.
In practice there are two problems with this approach. First, the propagation time difference is redshift-dependent (see Equation~\ref{eq:PhotonTimeDelay}), and only about 20\% of the well localized GRBs have an accurately measured redshift. 
The second problem is related to the reliability of the clock, i.e., to what extent neutrinos are emitted at the same time as, or close enough to, the burst of gamma rays. The evolution of GRBs pre, during and post the gamma-ray emission that is considered the time of the burst is still rather poorly known, and neutrinos could be emitted before, contemporaneously or after the gamma rays~\cite{Pitik:2021xhb}. Indeed the fireball model of GRBs~\cite{Piran:1999kx,Zhang:2003uk} predicts that neutrinos can be emitted in different energy ranges  during the precursor, burst and afterglow phases of a GRB. Searches for neutrino emission from GRBs with neutrino telescopes are therefore sometimes performed within a time window of up to several days around the detected time of the burst in gamma rays~\cite{IceCube:2021qyj}. It will be difficult to assign a difference between the arrival time of neutrinos and electromagnetic radiation from a single GRB to LIV effects if indeed neutrino emission from a GRB is detected in the future. There have been several proposals to try to ameliorate this problem by using similar populations of GRBs. Assuming that the production of neutrinos and photons is similar in similar GRBs, the intrinsic relative delay between neutrinos and photons due to production processes at the source should be independent of distance,  while it should depend on distance if the delay is due to LIV effects during propagation~\cite{Jacob:2006gn}. A similar experimental approach is to consider that if the arrival time correlation between gamma rays and neutrinos from GRBs follows a systematic pattern, e.g., neutrinos that can be associated with the position in the sky of an ensemble of similar GRBs, are detected systematically later or earlier than the electromagnetic radiation, that could indicate the effect of LIV~\cite{Amelino-Camelia:2015nqa}. However,  it would still be difficult to unambiguously allocate such effect to LIV, instead of a common production mechanism of neutrinos in GRBs. We need to advance our knowledge on the sequence of physical processes that lead and follow a GRB event (a relativistically expanding environment, still poorly known) in order to further distinguish with confidence new physics effects in such objects. Note also that these methods depend on an accurate measurement of the distance (redshift) of the GRBs.

 Even if the detection of an anomalous effect in the propagation of neutrinos would need a quite detailed knowledge of the time evolution of the source in order to unambiguously assign it to LIV effects, the non-observation of an unusual   propagation can be used to limit the LIV scale without such knowledge of the source. Indeed the only two extragalactic objects that have been identified as neutrino sources so far, the supernova SN1987A and the blazar TXS\,0506-056, have been used to such effect. The supernova SN1987A was the first celestial object to be detected in neutrinos, along with optical, X-ray and radio observations~\cite{Hillebrandt:1989uc}. The other object is the blazar TXS\,0506-056, which has been identified as a neutrino source as well as an optical, gamma ray and radio source~\cite{IceCube:2018dnn}. These two objects provide an actual example on how the ideas mentioned in the previous paragraph can be adapted to set strong limits on the scale of LIV effects. SN1987A was a supernova in the Large Magellanic Cloud, at a distance of 51.4 kpc, observed on February 23rd 1987. A total of 25 neutrinos were detected by three neutrino detectors in operation at the time (Kamiokande~\cite{Hirata:1988ad}, IMB~\cite{IMB:1987klg} and Baksan~\cite{Alekseev:1988gp}. See also~\cite{Aglietta:1987it}) during a short time interval of about 13 seconds, three hours before the visible light was detected. The total amount and energy of the detected neutrinos ($\cal{O}$(10) MeV) and the duration of the burst was found to be in accordance with the predictions of how a type-II supernova forms~\cite{Janka:2017vlw}.  A supernova explosion is a complex process where neutrinos are expected to be produced during different stages of the explosion through different process ($\beta$ decays of star material, $e^+$ and $e^-$ capture by nucleons or $e^+ e^-$ and $\nu \bar{\nu}$ annihilations) while they are thermalized through elastic scattering with nucleons, electrons and positrons. Although all neutrino flavours escape within a few seconds during the collapse, if one could follow the different phases of the explosion  with subsecond resolution, the development of the relative neutrino flux per flavour can be predicted as the collapsing star goes through distinct density phases. Different neutrino flavours feel different opacities in the medium due to their specific flavour-dependent interactions, and free streaming from the system occurs at different times. However none of the detectors mentioned above had such timing resolution, neither sensitivity to all neutrino flavours. The neutrinos detected from SN1987A were probably $\bar{\nu}_e$, through the process $\bar{\nu}_e + p \rightarrow n + e^+$, which has a dominant cross section over the process $\nu_e + e^- \rightarrow \nu_e + e^-$ at MeV energies (the Kamiokande collaboration estimated that only one neutrino among the 12 detected originated from the latter process). But neither a detailed picture of neutrino emission, nor of the relative time of neutrino emission with respect to photon release is necessary to set a limit on LIV using the neutrino signal from Kamiokande, IMB and Baksan. LIV effects on the propagation of neutrinos would result in an energy-dependent arrival time  (see Equation~(\ref{eq:PhotonTimeDelay}) 
 in the case of known $z_\mathrm{s}$), so the measured time span of the burst in the detectors, 13 seconds, can be used to set a limit on the strength of LIV effects, without any further knowledge of the neutrino emission sequence at the source. Crude as it is, this method gave at the time a much better limit on the scale of LIV than any obtained from neutrino beams in accelerators. If LIV effects on the propagation of neutrinos are parametrized at first order as $v/c=[1 \pm (E/E_\mathrm{QG})]$ (compare with Equation~\ref{eq:photonvelocity} for n=1), a limit on $E_\mathrm{QG}$ can be set requiring that the time spread in the arrival of the neutrinos is not larger than the observed window. This gives a 95\% confidence level limit $E_\mathrm{QG} > 2.7(2.5) \times 10^{10}$~GeV for subluminal(superluminal) neutrinos (and $E_\mathrm{QG} > 4.6(4.1) \times 10^{4}$~GeV if the dependence on energy is quadratic, i.e.,  $(E/E_\mathrm{QG})^2$)~\cite{Ellis:2008fc}. These are limits that are about five orders of magnitude more constraining than similar ones obtained from an analysis of MINOS data~\cite{MINOS:2007cyk}.

\begin{figure}[t]
    \includegraphics[width=\textwidth]{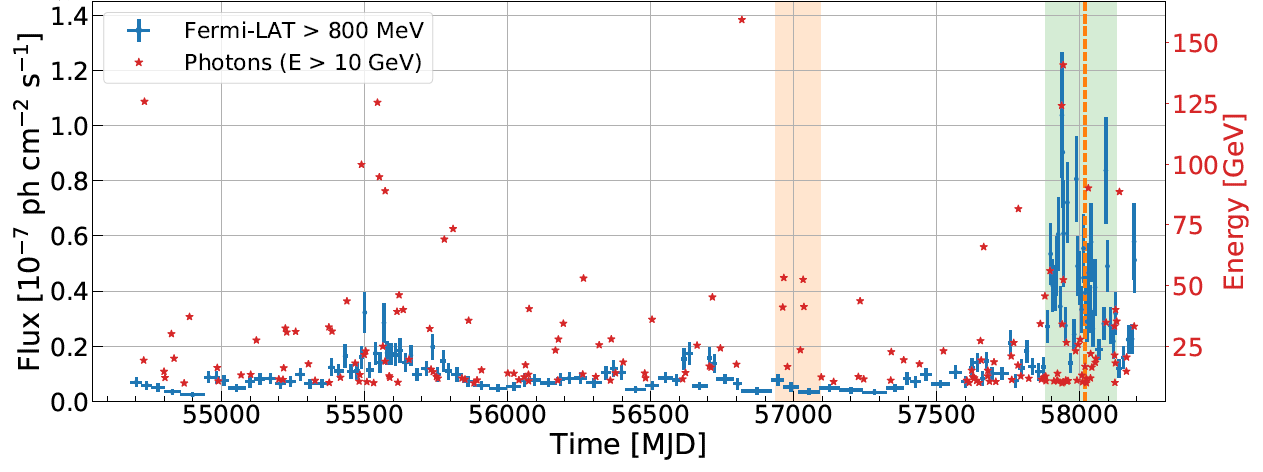}
  \caption{Gamma-ray flux of TXS\,0506+056 versus time, integrated above 800 MeV. The green vertical band denotes the gamma-ray flare of 2017/2018. The vertical red dashed line marks the IceCube neutrino alert IceCube-170922A and the vertical yellow band the neutrino flare detected by IceCube in archived data from 2014.  Figure from~\cite{Fermi-LAT:2019hte}. Reproduced by permission of the AAS.}
    \label{fig:txs-flare}
\end{figure}

The second object that has been used to extract limits on LIV is the blazar TXS\,0506-056. This blazar has a redshift $z = 0.3365\pm 0.0010$ and it is known to present flaring episodes, as Figure~\ref{fig:txs-flare} illustrates for the case of gamma rays. On September 22nd 2017 IceCube detected a neutrino with an estimated energy of 290~TeV from the direction of this object (vertical red dashed line in Figure~\ref{fig:txs-flare}) in coincidence with a gamma-ray flaring period of several weeks. Even if the neutrino emission time can not be assigned to a given specific gamma-ray flare to be able to assign a common clock between photons and neutrinos, the fact that neutrinos are detected within about 10 days of the photons can be already used to extract limits on the difference between their respective velocities in vacuum in a similar way as discussed above for SN1987A~\cite{Ellis:2018ogq,Wang:2020tej}. 
A difference in velocity between neutrinos and electromagnetic radiation induces a difference in arrival time of $\Delta t = \Delta v\,D$, where $D$ is the distance to the object. 
Assuming a LIV effect linearly proportional to energy, $\Delta v = -E/E_\mathrm{QG}$ where $E_\mathrm{QG}$ has the same interpretation as above, a limit of 10 days in $\Delta t$ translates into a limit of $E_\mathrm{QG} \gtrsim 3 \times 10^{16}$~GeV. This is still below the Plank mass, where quantum gravity Lorentz violating effects are assumed to surface. But it is a much more  restrictive  limit that the one obtained from SN1987A, illustrating the strength of using cosmological distances in these kind of studies.
 
 From the experimental point of view, there is a distinctive difference between using supernovae neutrinos with energy of a few ten MeV or high energy astrophysical neutrinos of TeV energies and above. MeV neutrinos do not provide a precise pointing in neutrino detectors and the association of a MeV neutrino burst with a supernova is done purely on timing.
 The electrons and positrons produced in the reactions $\nu_e + e^- \rightarrow \nu_e + e^-$ and  $\bar{\nu}_e + p \rightarrow n + e^+$ quickly scatter (and the positron annihilates), losing information from the direction of the incoming neutrino (the angular distribution of the SuperK and IMB SN1983A events is practically isotropic with respect to the position of the source). For larger, km$^3$-size, neutrino telescopes the situation is not better due to the large distance between optical sensors, up to $\cal{O}$(100)~m, in comparison with the short electron (and positron) tracks of less than a meter length (the mean free path of a 10 MeV electron in water/ice is about 5.6 cm~\cite{Kopke:2017req}). 
 However, the copious neutrino flux expected from a supernova explosion (between $10^{57}-10^{58}$ neutrinos in total) induces an increase of events in the detectors during a few seconds, which constitutes the signal. In the case of neutrino telescopes a supernova signal would consist of a coherent increase in optical sensor noise across the whole array for the few seconds of the burst, but without the possibility of identifying individual electrons or positrons from the neutrino interactions, neither with the possibility of pointing to the source, e.g.~\cite{Kopke:2017req}. 
  
 On the other hand, TeV-PeV neutrinos accompanying GRBs or AGN flares are extremely rare and good pointing is needed to associate the neutrino with the object, which is only achieved with the long muon tracks produced in charged-current $\nu_{\mu}$ interactions. Therefore, of the three neutrino flavours, only one is usable in practice if a precise association with the position of the source is required. The advantage is that the expected background from atmospheric neutrinos from the same direction and within the chosen time window can be negligible, less than 0.01 events in a kilometer cube detector during the time of a few seconds burst, or a few events if the time window is of the order of several days (of course larger time windows allow for a higher probability of background sneaking in, weakening the usefulness of the method). 

 There is another way of using multimessenger probes to study LIV effects without the need of relative timing by taking into account the fact that LIV allows processes like ``vacuum neutrino bremsstrahlung''\footnote{This process is usually known as vacuum Cherenkov emission although it does not really resemble Cherenkov radiation since the usual Cherenkov radiation is emitted by the media which the relativistic particle traverses, not by direct radiation from the particle itself.}, $\nu \rightarrow \nu \gamma$~\cite{Jacobson:2002hd}. In this scenario, the astrophysical neutrino flux would be a source of gamma rays, contributing to the total diffuse flux that has been measured by different experiments, e.g.~\cite{Fermi-LAT:2018udj}. By requiring that neutrino bremsstrahlung due to LIV should not contribute to the diffuse gamma-ray flux in a way that contradicts the measurement, a limit on the strength of LIV can be established. Another consequence of the above mentioned process, along with $\nu \rightarrow \nu e^+e^-$, which is also allowed in LIV models~\cite{Somogyi:2019yis}, would be a new ``GZK-like'' effect (see Section~\ref{subsec:CRs}) for high energy neutrinos. Astrophysical neutrinos would lose energy during their propagation, appearing at lower energies in the spectrum and showing a cutoff at an energy scale that depends on the LIV strength. The observation of astrophysical neutrinos up to a given energy can thus be used to set limits on  LIV~\cite{Christian:2004xb,Mattingly:2009jf,Borriello:2013ala,Stecker:2014oxa,Stecker:2014xja,Tomar:2015fha}. These two methods rely on a precise measurement of the astrophysical neutrino flux and knowledge of whether it presents a cutoff or not (and at what energy if that is the case), something that is not established at this point. Still, the sheer observation of the first PeV neutrinos by IceCube, implying that any cutoff in the neutrino spectrum lies above such energy, allowed to improve limits on several LIV-coefficients of the SME by up to 20 orders of magnitude (!) with respect to previous limits at the time~\cite{Diaz:2013wia}.

\paragraph{\textit{ Flavour interferometry}}
This technique makes use of the two classes of ``beam'' that a neutrino telescope is subject to: the copious, but relatively short-baseline (at most Earth-diameter), atmospheric neutrino flux or the much weaker, but cosmological-baseline, astrophysical flux. Both have their advantages and disadvantages.

The advantage of using atmospheric neutrinos to address new physics with neutrino telescopes is that the flux is relatively well known.  ``Conventional'' atmospheric neutrinos arise mainly from the decay of pions and kaons produced in cosmic ray interactions and their spectrum follows a power-law as $\Phi(E)\propto E^{-\gamma}$, with $\gamma$ typically quoted as $-3.7$, although the actual index depends on the energy range under consideration. This flux has been measured by several experiments to a good agreement with theoretical calculations~\cite{Gaisser:2019efm}. A ``prompt'' component arises from the decay of heavier D$_{\mathrm s}$ and B$_{\mathrm s}$ mesons containing a charm or bottom quark. This prompt flux can become comparable to the conventional atmospheric neutrino spectrum at energies above several hundred TeV. Incidentally, this is the energy where the astrophysical neutrino flux becomes also stronger than the conventional atmospheric neutrino flux, so the prompt flux remains undetected so far. IceCube measurements of the neutrino flux above 100~TeV provide only upper limits on the contribution of the prompt flux, with a lower limit compatible with zero~\cite{IceCube:2020wum}.

Without entering into the known details of neutrino oscillation theory (see e.g.~\cite{Fantini:2018itu} for a review) let us just mention a few facts that will be useful to understand how they can be used to detect LIV effects. Assuming a two-flavour scenario for illustration purposes, a neutrino state of flavour $\alpha$, ${\ket {\nu_{\alpha}}}$, can be expressed as a superposition  of the mass states $1$ and $2$, with masses $m_1$ and $m_2$ respectively,
\begin{equation}
 {\ket {\nu_{\alpha}(t)}}\,=\, -\mathrm{sin}\theta e^{-iH_1 t}{\ket {\nu_{1}}} + \mathrm{cos}\theta e^{-iH_2 t}{\ket {\nu_{2}}}   \; ,
  \label{eq:kett1}
\end{equation}
where the Hamiltonian for free-propagating neutrinos, $H_{1,2}$, is based on the dispersion relation  $E^2_i=p_i^2 +m_i^2$ ($i=1,2$) and $\theta$ is the "mixing angle", used to parametrize the mass composition of the flavour state while keeping unitarity. The neutrino flavour at the detector, ${\ket {\nu_{\alpha}}}$, is related to the flavour at production, ${\ket {\nu_{\beta}}}$, by the known relation 

\begin{figure}[t]
  \begin{minipage}[c]{0.40\textwidth}
    \includegraphics[width=\textwidth]{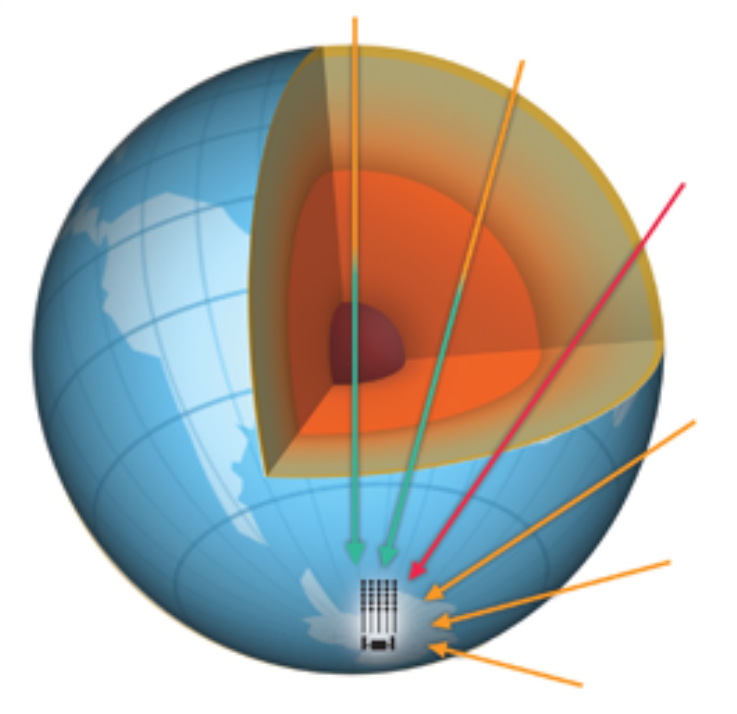}
  \end{minipage}\hfill
  \begin{minipage}[c]{0.50\textwidth}
    \caption{Measuring neutrino oscillations with atmospheric neutrinos. Since the flavour oscillation probability depends on the path-length from production to detector, a measurement of such path-length for atmospheric neutrinos is the zenith angle they arrive to the detector. A measurement of zenith angle and energy shows the typical flavour oscillation pattern, as shown in Figure~\ref{fig:atm_osc}. Illustration courtesy of IceCube collaboration.}
    \label{fig:earth}
  \end{minipage}
\end{figure}
\begin{figure}[t]
  \begin{minipage}[c]{0.50\textwidth}
    \includegraphics[width=\textwidth]{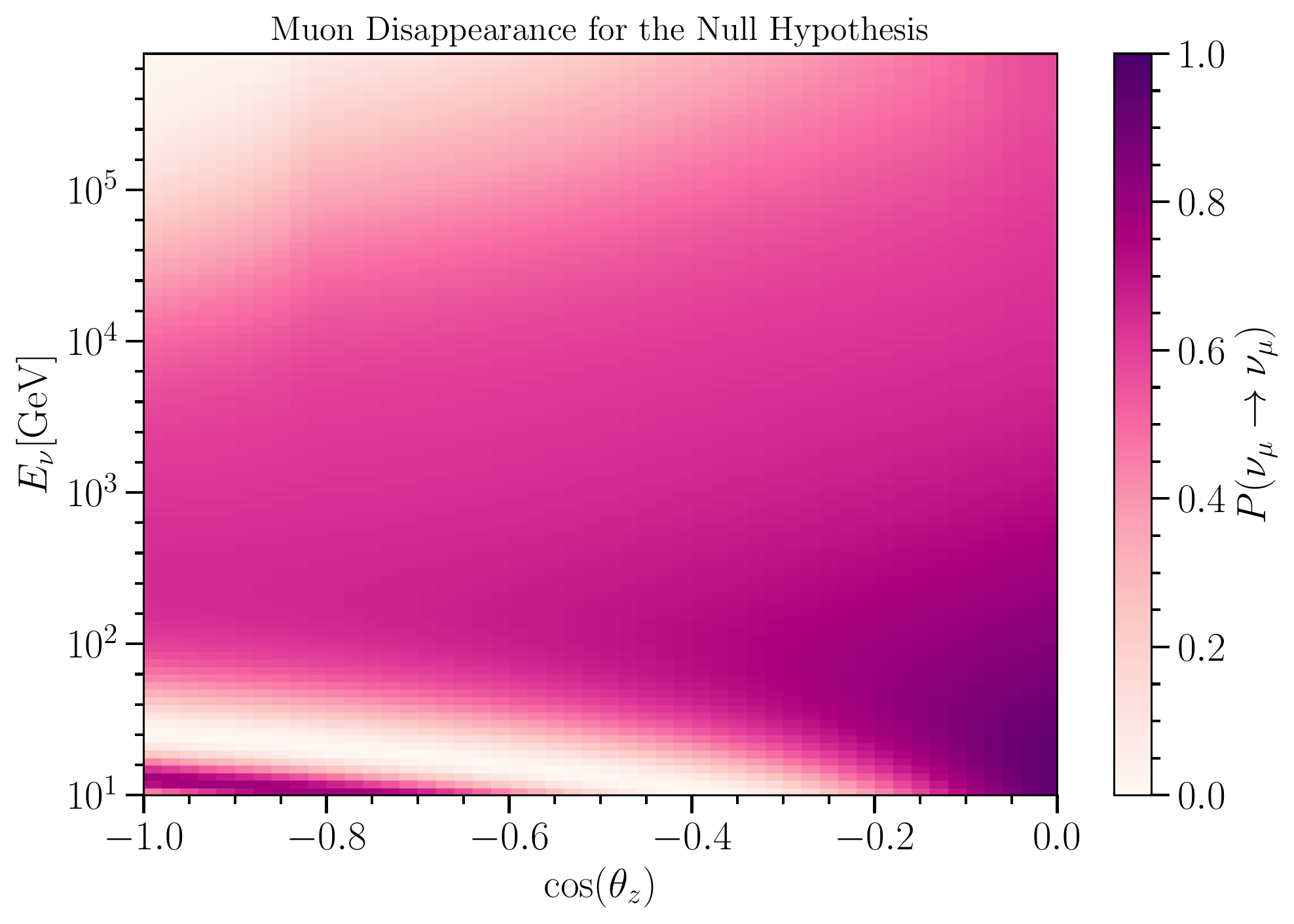}
  \end{minipage}\hfill
  \begin{minipage}[c]{0.50\textwidth}
    \includegraphics[width=\textwidth]{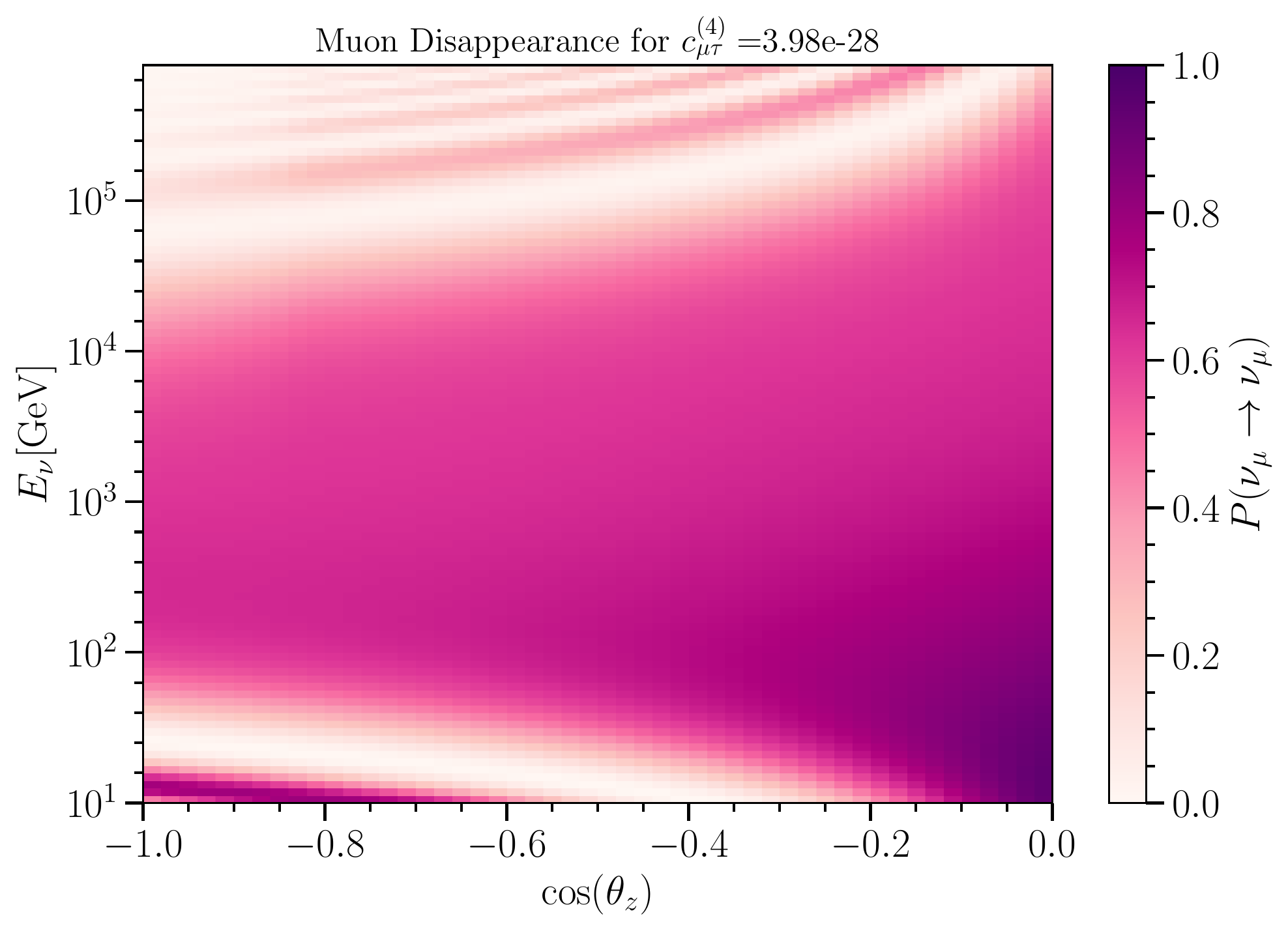}
  \end{minipage}
    \caption{$\nu_{\mu} \rightarrow \nu_{\mu}$ survival probability, denoted as the color code, as a function of energy and arrival direction between vertically up-going ($\mathrm{cos}\theta_z = -1$) and horizontal ($\mathrm{cos}\theta_z =0$). {\bf Left:} Standard oscillation scenario. {\bf Right:} Oscillations under the assumption of a Lorentz-violating term proportional to energy with strength set to the current limit. Plots courtesy of B. Skrzypek.}
    \label{fig:atm_osc}
\end{figure}


\begin{equation}
 {\ket {\nu_{\alpha}}}\,=\, \sum_{\beta=\mathrm{e},\mu,\tau} P\left(\nu_{\beta} \rightarrow \nu_{\alpha} \right)  {\ket {\nu_{\beta}}}_{\mathrm{S}}\; ,
\end{equation}
where S stands for ``source'' and the transition probability $ P\left(\nu_{\beta} \rightarrow \nu_{\alpha} \right)$ represents the standard flavor oscillations in the absence of new physics,
which is linearly proportional to the square mass difference of the mass eigenstates and the propagation length, and inversely proportional to the neutrino energy,
\begin{equation}
  P\left(\nu_{\beta} \rightarrow \nu_{\alpha} \right) = \mathrm{sin}^2 2\theta \mathrm{sin}^2 \left( \frac{\Delta m_{1,2}^2 L}{4 E} \right)
  \label{eq:std_osc}
\end{equation}
where $\Delta m_{1,2}^2 = m_2^2 - m_1^2$, measured in $eV^2$, $E$ is measured in GeV and $L$ in km, and where it has been assumed that neutrinos are relativistic and one can set $t \sim L$ in Equation~(\ref{eq:kett1}), the distance traveled between production and detection. 
Figure~\ref{fig:earth} illustrates the idea behind using atmospheric neutrinos to measure neutrino oscillations. If the neutrino energy can be measured to a suitable precision, and it can in neutrino telescopes, then the path-length from production to the detector is related to the arrival direction of the neutrino, which in local detector coordinates is the zenith angle. So a measurement of $L/E$ is possible in Equation~(\ref{eq:std_osc}). The left plot of Figure~\ref{fig:atm_osc} shows the expected atmospheric $\nu_\mu$ disappearance probability under the assumption of standard oscillations as a function of neutrino energy and arrival direction.

As a simple illustration on how anomalous flavour-changing effects due to Lorentz invariance violation can modify the energy and zenith angle distributions of atmospheric neutrinos let us consider the following simple example.  As mentioned above, one of the effects of LIV can be the modification of the dispersion relation for massive particles (Equation~\ref{eq:generalMDR}). Let us assume here for generality that the parameter $\eta$ is eigenstate dependent, $\eta_i$ (here $\eta$ can still be positive or negative, but is not restricted to values $\pm1$ any more). Under this assumption, the Hamiltonian in Equation~(\ref{eq:kett1}) is different from the free-propagation case, and a similar calculation that yields Equation~(\ref{eq:std_osc}) now gives (e.g.~\cite{Morgan:2008fof}),

\begin{equation}
P\left(\nu_{\beta} \rightarrow \nu_{\alpha} \right) = sin^2 2\theta sin^2 \left( \frac{\Delta m_{1,2}^2 L}{4E} + \frac{\Delta\eta E^{n +1} L}{4 E^{n}_{\mathrm{P}}} \right)
  \label{eq:osc_prob2}
\end{equation}
where  $\Delta\eta = \eta_2 - \eta_1$. If $\Delta\eta = 0$, i.e., there is no difference in the propagation of the two neutrino mass eigenstates due to LIV effects, we recover the standard oscillation formula of Equation~(\ref{eq:std_osc}). Note that in this scenario neutrino oscillations are only sensitive to differences in the strength of LIV effects on mass eigenstates, and not on the individual $\eta$s, in a similar way that standard oscillations are sensitive to the difference of the masses squared and not to individual masses. The previous parameterization of LIV effects on neutrino propagation provides a useful way to perform experimental searches, since the parameters $\Delta\eta$ and $n$ can be probed by measuring $L$ and $E$ in neutrino telescopes and comparing the result to the expected oscillation pattern without LIV. The most common way to extract limits on LIV effects is by using oscillograms, two-dimensional plots of the oscillation probability versus $E$ and $\mathrm{cos}(\theta_z)$, as the ones shown in Figure~\ref{fig:atm_osc}. By comparing the data with the expected pattern of a LIV model, the parameters of the model in question can be constrained~\cite{Super-Kamiokande:2014exs,Morgan:2008fof}. Note that in the example shown in the right plot of Figure~\ref{fig:atm_osc} the effects of LIV appear at high energies, where neutrino telescopes are specially sensitive with respect to smaller neutrino detectors or accelerator experiments. Indeed Super-Kamiokande data were readily used already in 1999 to search for an anomalous oscillation pattern due to new physics, among it LIV, and limits set at the level of $|1-\beta_{\nu}| < 10^{-24}$, where $\beta_{\nu}$ represents the neutrino velocity in units of $c$~\cite{Fogli:1999fs}.

\begin{figure}[t]
  \begin{minipage}[c]{0.47\textwidth}
    \includegraphics[width=\textwidth]{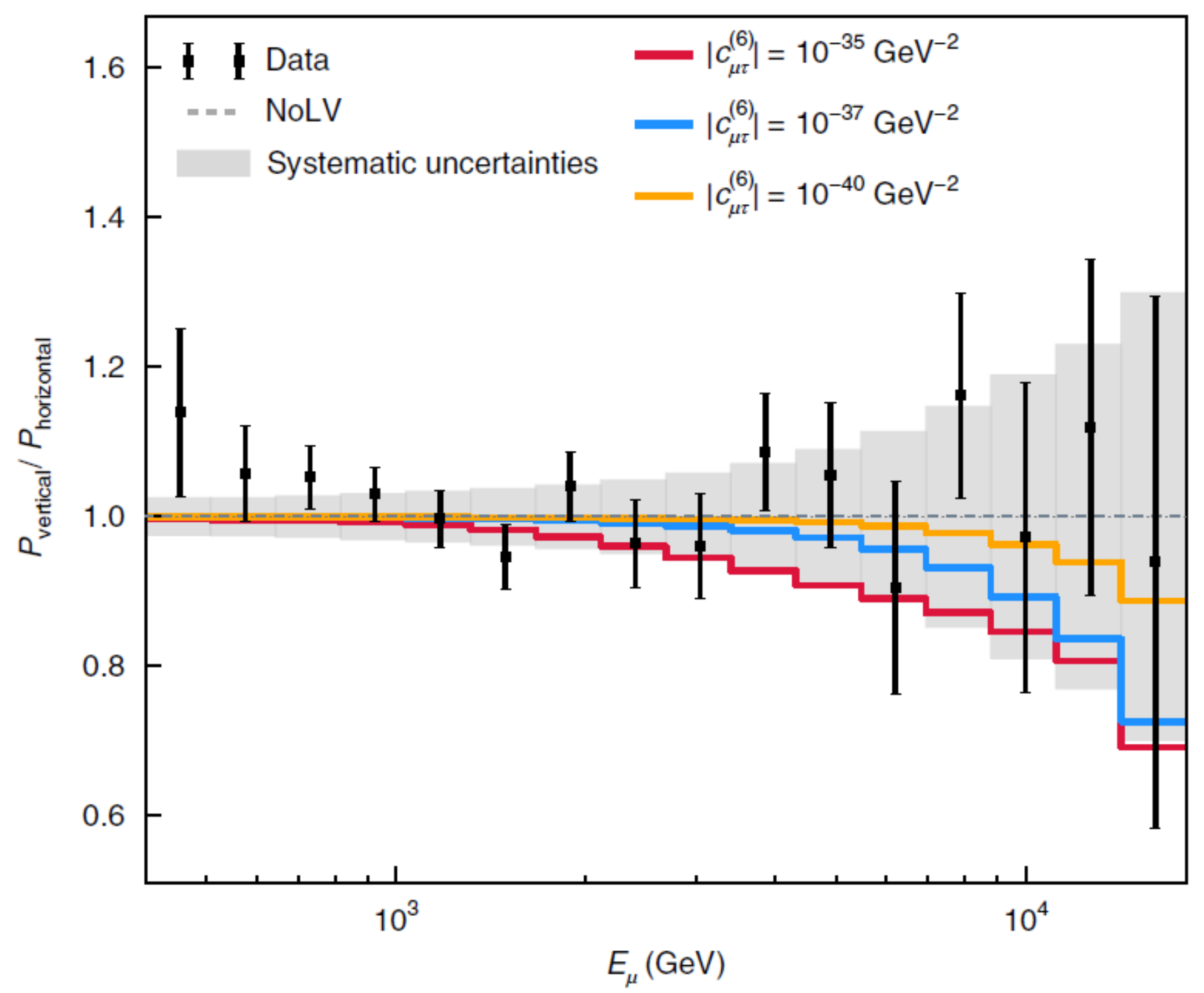}
  \end{minipage}\hfill
  \begin{minipage}[c]{0.50\textwidth}
    \includegraphics[width=\textwidth]{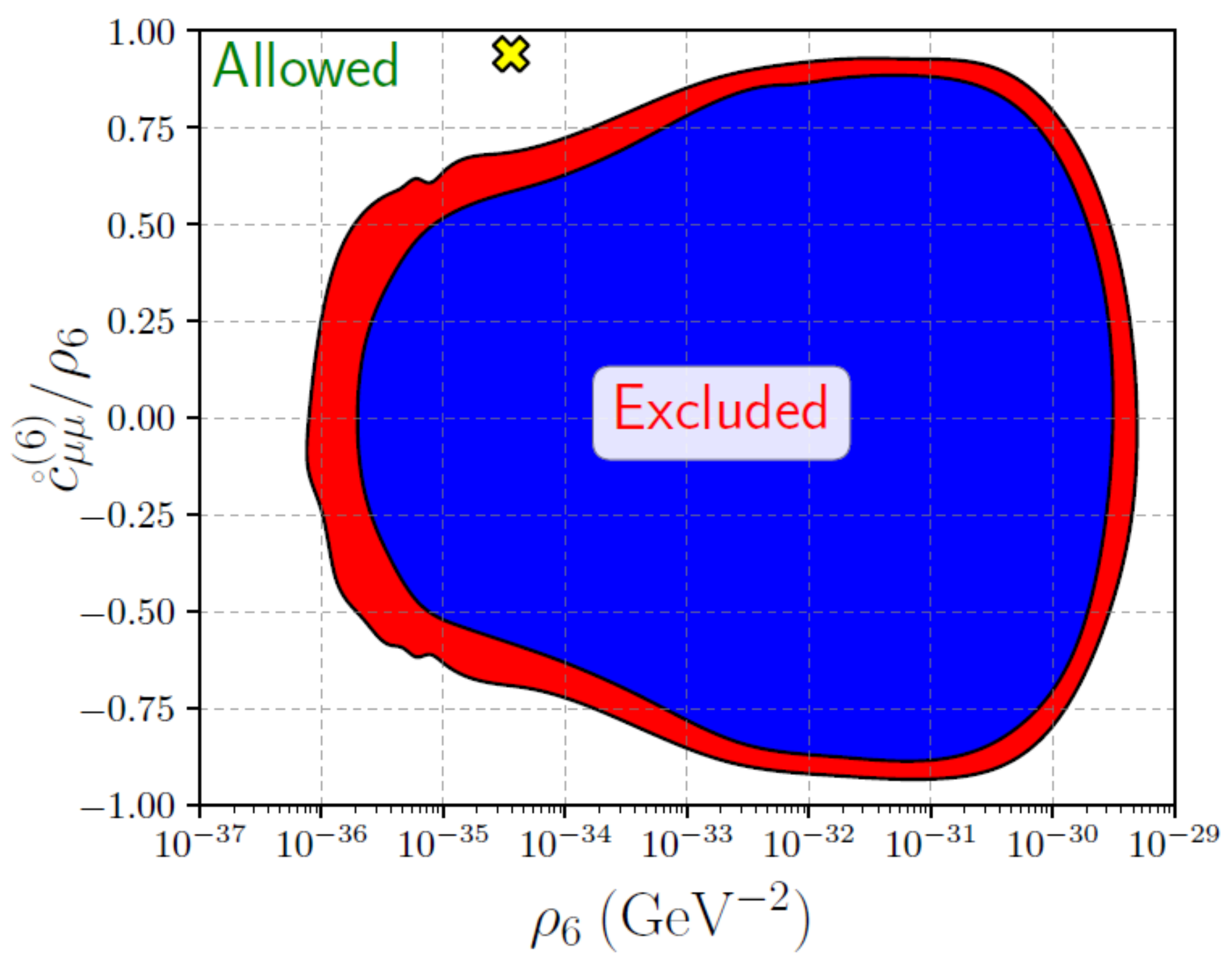}
  \end{minipage}
    \caption{{\bf Left:} Ratio of the arrival probability of vertical to horizontal events in IceCube, as a function of energy. The coloured lines show the prediction for different strength of LIV effects, defined as different values of the relevant parameters in the SME Hamiltonian. The null hypothesis (expected event ratio under standard oscillations) is shown as the dashed, straight horizontal line. Data is shown as the black dots with accompanying statistical error bars.  {\bf Right:} Example of the excluded parameter space for one of the SME coefficients responsible for LIV. The parameter $c_{\mu\mu}^{(6)}$ represents the contribution to Lorentz violation that can be extracted from muon disappearance measurements. The parameter $\rho_6$ represents the total strength of the Lorentz violation effect. The super/sub-script 6 indicates the dimension of the operator in the SME Hamiltonian. The best-fit point is shown by the yellow cross and the blue (red) region is excluded at 99\%
(90\%) confidence level respectively. Both figures from~\cite{IceCube:2017qyp}. }
    \label{fig:atm_flavour}
\end{figure}

A rather straightforward way to search for LIV effects with atmospheric neutrinos is to exploit the fact that the relatively short distance of travel for horizontal neutrinos ($\mathrm{cos}(\theta_z)=0$) is not enough to develop any appreciable spectral distortion due to LIV, even at high energies. However, the effect becomes maximal for up-going neutrinos crossing the Earth ($\mathrm{cos}(\theta_z)=-1.0$), as can be seen on the right plot of Figure~\ref{fig:atm_osc}. So the ratio of transition probabilities of vertical events to horizontal events can be used to determine, or set limits to, LIV parameters within a given model. Indeed the Super-Kamiokande, ANTARES and IceCube collaborations have performed searches for LIV using this technique~\cite{Super-Kamiokande:2014exs,Morgan:2008fof,IceCube:2017qyp}. The left plot of Figure~\ref{fig:atm_flavour} shows such ratio from a search for LIV effects with atmospheric neutrinos by the IceCube collaboration. An example of how specific parameters of the Lorentz invariance terms in the Lagrangian can be constrained is shown in the right plot of  Figure~\ref{fig:atm_flavour}, with more examples available in the additional material of reference~\cite{IceCube:2017qyp}.  This is a simple and powerful approach to access LIV effects in the neutrino sector, thanks to the high sensitivity that the oscillation patterns shows to deviations from the standard scenario. 

An intriguing possibility of breaking Lorentz invariance is the introduction of an anisotropy in the space-time (equivalently, a preferred spatial direction) due to the coupling of the Lorentz violation terms with the neutrino propagation vector $\vec{p}$. This will induce an oscillation probability that is dependent on the arrival direction of the atmospheric neutrinos which, since the detector is fixed to the Earth, means a time-dependent event rate at the detector in the form of a sinusoidal neutrino disappearance signal. Searches for such an effect have been performed with IceCube data by measuring the event rate as a function of right ascension and, even with a limited data sample~\cite{IceCube:2010fyu}, competitive constraints were set to the coefficients of the SME Hamiltonian responsible for the predicted anisotropy. This is an effect that can also be searched for with accelerator neutrino beams at a complementary energy regime as has been reported in~\cite{MiniBooNE:2011pix,T2K:2017ega,MINOS:2012ozn}. An equivalent analysis was performed by the SNO collaboration, but using solar neutrinos~\cite{SNO:2018mge}. In this case the pathlength is the radius of the orbit of the Earth and the expectation is a modulation on the oscillation probability of electron neutrinos over the course of a year, as the Earth moves in the frame of the Sun. The results of the SNO analysis set limits for the first time on 38 previously unconstrained parameters of the SME and confirmed that LIV effects do not occur below an energy scale of about $10^{17}$~GeV. 

\begin{figure}[t]
  \begin{minipage}[c]{0.45\textwidth}
    \includegraphics[width=\textwidth]{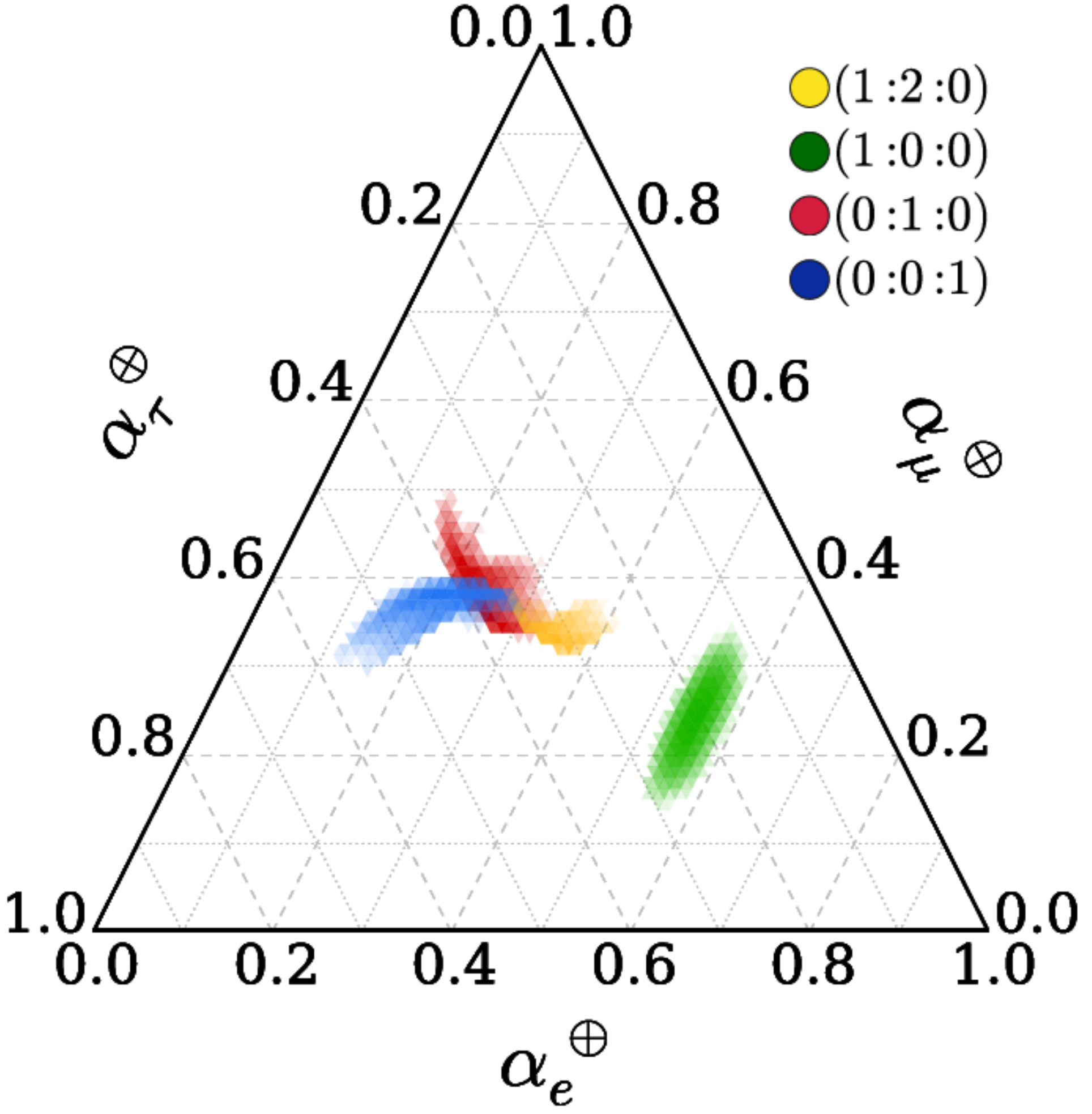}
  \end{minipage}\hfill
  \begin{minipage}[c]{0.50\textwidth}
    \includegraphics[width=\textwidth]{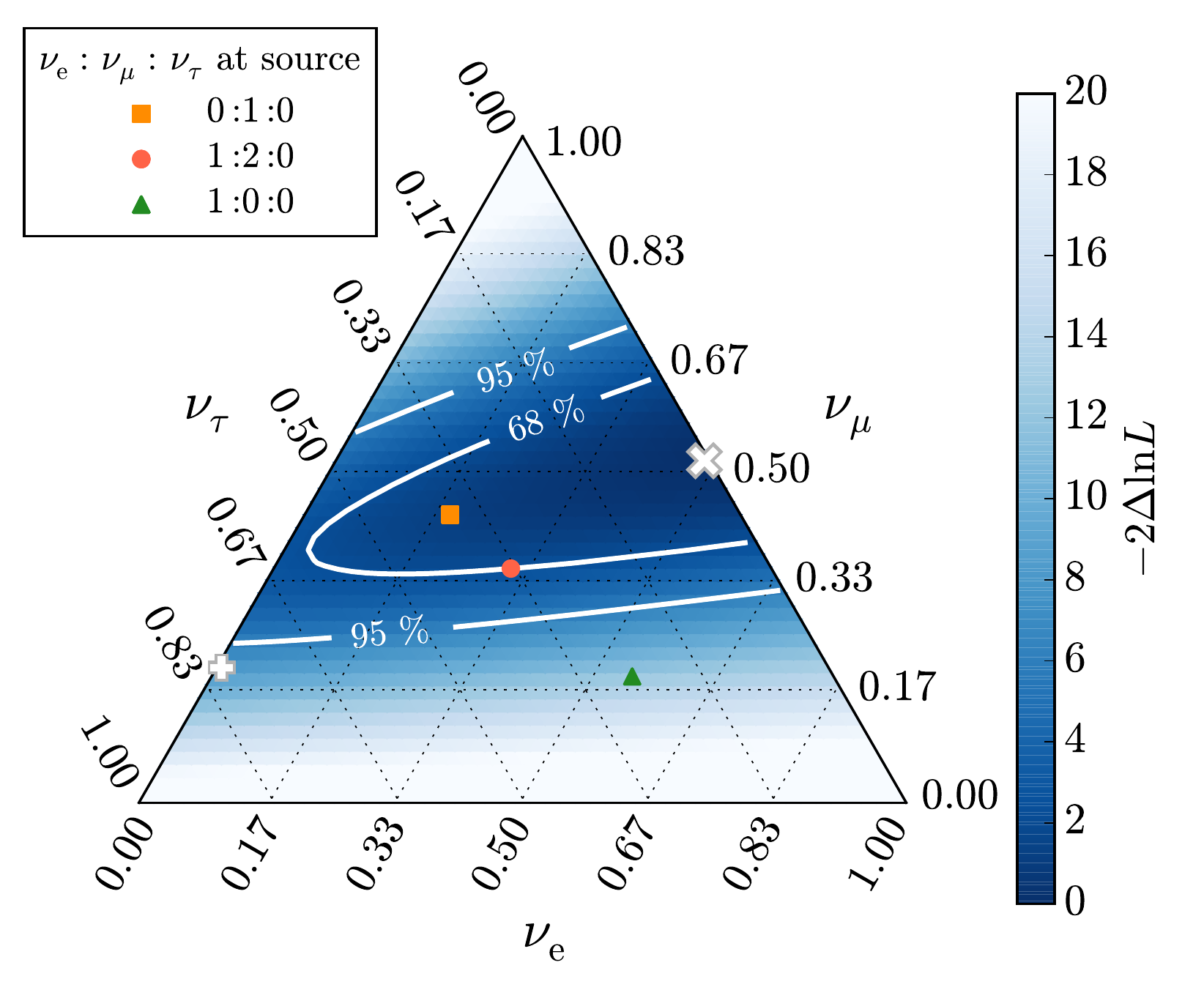}
  \end{minipage}
    \caption{{\bf Left:} Allowed neutrino flavour composition at Earth (colored areas) when different flavour compositions at the source are assumed (corresponding colour point). The original flavour composition is washed out by standard oscillations over cosmological distances. The three axes of the plot show the fraction of each neutrino flavour, with pure $\nu_{e}$, pure $\nu_{\mu}$ and pure $\nu_{\tau}$ at the respective vertices. Current neutrino oscillation parameters have been used. Any measured flavour ratio of astrophysical neutrinos outside the coloured areas would point to new physics, rather independently of any source composition. Figure from~\cite{Arguelles:2015dca}. {\bf Right:} Current measurement of the flavour triangle with astrophysical neutrinos by the IceCube collaboration. The best-fit point is marked with a white cross (``x'') and the 68\% and 95\% contours are indicated with the white lines. The orange, red and green points mark the expected composition at the detector from the assumptions at source given in the upper-left caption of the figure with the same color. Although a neutron-rich source composition is disfavoured, the results are compatible with standard oscillations and a source flavour composition of both ${\nu_e:\nu_{\mu}:\nu_{\tau}} = {1:2:0}$ or ${\nu_e:\nu_{\mu}:\nu_{\tau}} = {0:1:0}$. Figure from~\cite{IceCube:2015gsk}.}
    \label{fig:osc_triangle}
\end{figure}

However, the most sensitive search for LIV effects is achieved with astrophysical neutrinos. The extremely long path-lengths can compensate the weak expected effects and the results of searches can probe up to the Planck regime. In principle this approach relies on an assumption about the original neutrino flavour ratio at the source, something unknown, and on the shape of the neutrino energy spectrum, something currently measured with limited precision~\cite{IceCube:2020wum}. If neutrinos in astrophysical sources are produced in pion and kaon decays, the same flavour ratio at the source is expected as in atmospheric neutrinos, ${\nu_e:\nu_{\mu}:\nu_{\tau}} = {1:2:0}$. But that might not be the case in scenarios with strong magnetic fields, where muons can lose a significant fraction of their energy before decaying, and the high energy neutrinos produced arise only from the direct pion decay. In this case the expected flavour composition at the source is ${\nu_e:\nu_{\mu}:\nu_{\tau}} = {0:1:0}$~\cite{Hummer:2010ai}. On the other hand, neutron-rich sources can emit neutrinos through neutron decay producing a flavour composition at the source of ${\nu_e:\nu_{\mu}:\nu_{\tau}} = {1:0:0}$ (although this scenario seems to be disfavoured by current IceCube data~\cite{Bustamante:2019sdb}). It would seem therefore that an exact knowledge of the initial flavour composition at the source is key to be able to study LIV effects arising during cosmological propagation of neutrinos. 
In practice, however, the dependence on the source flavour composition is ameliorated since the flavour composition at Earth, after the neutrinos travel over cosmological distances, is reduced to a small region around ${\nu_e:\nu_{\mu}:\nu_{\tau}} = {1:1:1}$ in the flavour parameter space, practically independent of the original flavour composition at the source.  The left plot of Figure~\ref{fig:osc_triangle} illustrates this effect~\cite{Arguelles:2015dca}. The three axes of the plot show the fraction $\alpha_{(e,\mu,\tau)}$ of each neutrino flavour, with pure $\nu_{e}$, pure $\nu_{\mu}$ and pure $\nu_{\tau}$ at the respective vertices. The fraction of each flavour at any point in the triangle is obtained by projecting the point onto each axis in the order ($e$, $\mu$, $\tau$) and in the direction that points closer to the origin of each axis, so that the sum of the projections is one. The coloured regions show the allowed flavour composition at Earth starting from a flavour composition at the source indicated by the color code, assuming standard oscillations and no new physics. Although there is some ``memory'' of the source flavour ratio, the allowed flavour compositions at Earth concentrate around well delimited regions. A measured flavour ratio outside these regions points to new physics. This is a very powerful method to access new physics effects, not only due to LIV~\cite{Song:2020nfh}, in a way that does not depend on relative arrival timing between different messengers and it is quite independent of the neutrino production process at origin. The right plot of Figure~\ref{fig:osc_triangle} shows results from the IceCube collaboration obtained with the high-energy astrophysical neutrino flux~\cite{IceCube:2015gsk}.  The white cross in the plot marks the best-fit point from the analysis and the  allowed 68\% and 95\% confidence level regions on the flavour composition of astrophysical neutrinos are marked with the white lines. The plot shows that current data is still compatible with the expectation of standard oscillations and typical source flavour compositions, marked by the  orange, red and green points in the upper-left caption of the figure with the same color. More data is needed to reduce the allowed regions and make a more precise measurement of the flavour ratio to be able to establish if any deviation from the null hypothesis (no LIV) is present.

\subsection{Cosmic rays}
\label{subsec:CRs}
\begin{figure}[t]
		\centering\includegraphics[width=0.6\textwidth]{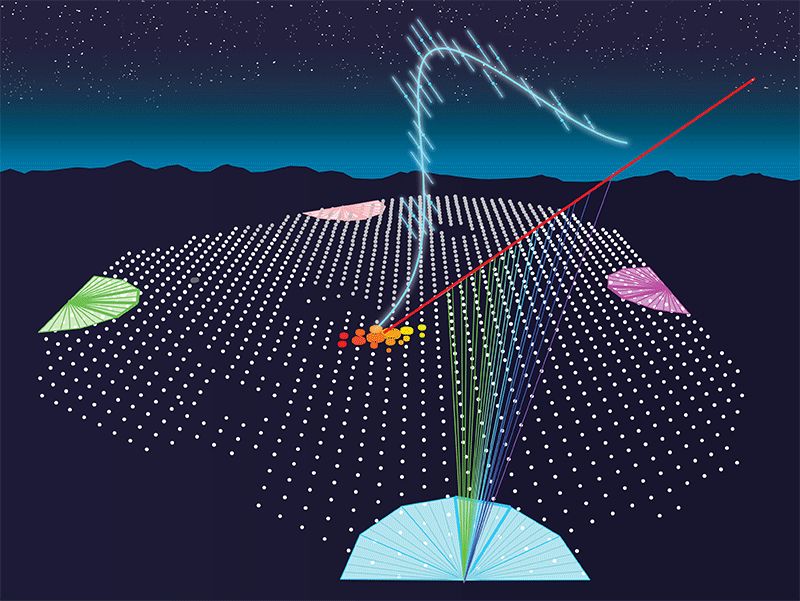}
	\caption{Illustration of the hybrid detection of an air shower in the Pierre Auger Observatory. The white dots represent the surface water Cherenkov tanks while the semicircles show the location and field of view of the fluorescence telescopes. The red line represents the original direction of the primary cosmic ray. The blue curve shows the shower particle density as a function of atmospheric depth, directly accessible through the measurement of fluorescence light by the four telescopes. Muons and electrons that reach the surface trigger several of the water Cherenkov tanks, represented by the coloured dots. Figure reproduced with permission by APS/Carin Cain,  Physics 9, 125 (2016)}
	\label{fig:auger_event}
\end{figure}

 While any particle arriving at the Earth from outer space can be considered a cosmic ray, in this section we will just focus on protons and heavier nuclei of ultrahigh energies ($10^{18}$ eV and above). 
 
The interaction of a cosmic ray with an atom in the higher atmosphere, typically occurring at an altitude of about 25-30~km, produces a cascade of particles that develops through the atmosphere reaching a maximum particle density at a depth that depends on the interaction energy. Stable particles can reach the ground over an area that can cover several km$^2$.  The direction and energy of the primary cosmic ray are reconstructed with the help of large surface arrays of particle detectors\footnote{Cosmic rays can also be detected directly by placing particle detectors in space, e.g.~\cite{Boezio:2020pyu,Battiston:2020loc}, although the necessarily small detection area of these kind of detectors limit their energy reach and detection rate.}. Due to the impossibility to tightly instrument such large surface areas, air shower arrays are segmented detectors, consisting of ``stations'' separated by a given distance and specialized in detecting electrons and muons. We already discussed water Cherenkov tanks and/or scintillators in Section~\ref{subsec:GRs}. They are commonly used to detect the electromagnetic component of the shower at ground level, but the array can be supplemented with telescopes that detect the fluorescence light produced by the excitation of nitrogen atoms in the atmosphere during the development of the particle shower, see Figure~\ref{fig:auger_event}. The largest air shower arrays in operation are the Pierre Auger Observatory in Argentina~\cite{PierreAuger:01, PierreAuger:2015eyc}, Large High Altitude Air Shower Observatory (LHAASO) in Tibet~\cite{Ma:2022aau} and the Telescope Array in Utah~\cite{TelescopeArray:01, TelescopeArray:2008toq}, with primary energy thresholds of $10^{17}$~eV, $10^{12}$~eV and $10^{16.5}$~eV respectively. The development and ground footprint of an air shower started by a high energetic gamma, a proton or heavy nucleus are slightly different, and they can be distinguished. In this section we discuss only the cosmic ray detection capabilities of air shower arrays since photon detection has been discussed in section~\ref{subsec:GRs}.

 The original direction of the cosmic ray primary can be obtained by measuring the relative arrival time to each detector unit of the particles that reach the ground, while the amount of particles detected at the ground is a proxy for the primary energy. Both the measurement of primary direction and energy is aided by the measurement of the fluorescence light during the development of the shower in the atmosphere. Note that the estimation of the energy of the primary relies in detailed Monte Carlo simulations of particle production and decay during the development of the  particle shower, which in turn relies on physical quantities like cross sections that have been measured in the laboratory, but not at the center of mass energies reached in cosmic ray interactions (the threshold energy of the Pierre Auger array in the CM frame, a $10^{17}$~eV proton hitting a stationary nucleon in the atmosphere, is comparable with the 14~TeV CM energy at the LHC). Particle production in air showers also involve processes that can not be perturbatively calculated in QCD, so there is a degree of model dependency in the translation of measured particle multiplicity at ground level and primary energy. 
 
The main difference between cosmic rays and the messengers considered in the previous sections is that cosmic rays are electrically charged and their propagation over astrophysical distances is affected by intergalactic and galactic magnetic fields. Given the poorly known structure and distribution of intergalactic magnetic fields, it is difficult to precisely know above which energy a cosmic ray can point to its source, but it seems clear that one needs energies above $10^{18}$~eV to achieve pointing of order a degree resolution~\cite{AlvesBatista:2018kup}. This means that cosmic ray sources can not be located as precisely as in the case of gammas or neutrinos, neither can cosmic rays be used in coincidence with electromagnetic or neutrino emission to measure relative arrival times.

However, there are still tests of LIV that can be performed with the observed cosmic ray flux without the need to identify the sources. And there are two levels at which this can be done: either by looking into LIV effects during the propagation of the cosmic rays, or by looking for LIV effects in the development of the particle shower once the cosmic ray has interacted in the atmosphere. Note that the predictions on how a particle shower evolves (production cross sections of different particle species, lifetimes and particle propagation and energy losses in a medium) are based on Lorentz invariant Standard Model processes. Modified particle kinematics due to LIV will modify the expected relative particle composition and spectrum of air showers at the surface, so any deviation from the expected signal in an air shower detector can be interpreted as new physics. This is easier said than done, since the underlying assumption is then that the development of a particle shower is precisely understood in terms of standard processes and their uncertainties, which is currently far from obvious at energies beyond those available at accelerators or beyond the regime where perturbative QCD can be used.

\begin{figure}[t]
	\centering\includegraphics[width=0.8\textwidth]{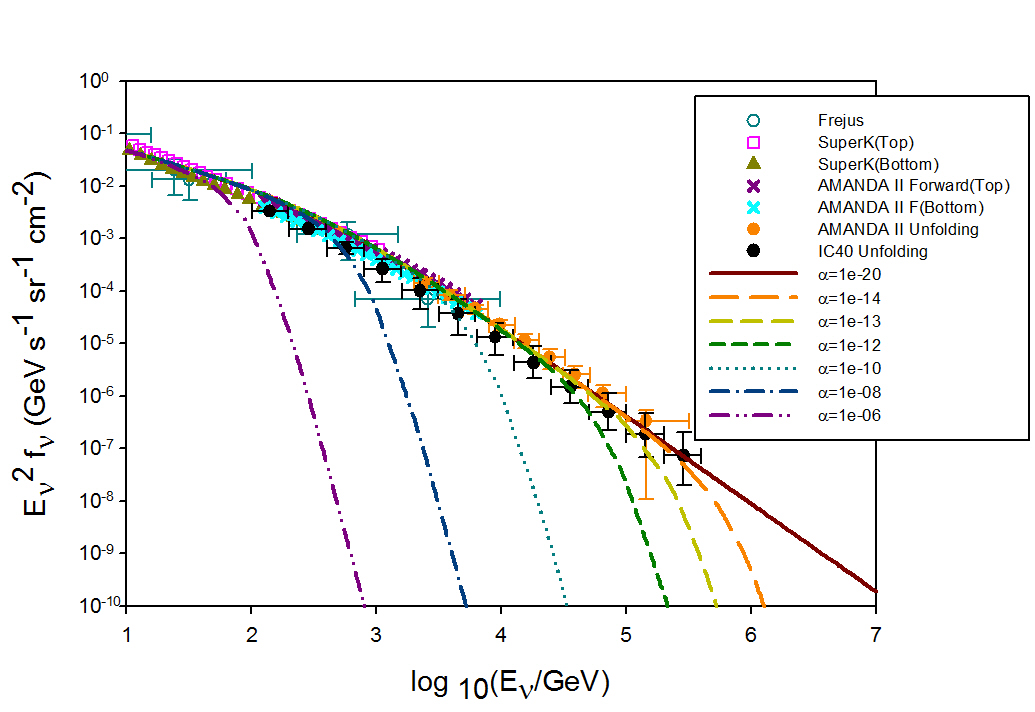}
	\caption{Expected atmospheric muon neutrino energy spectrum for different values of LIV strength, denoted by the parameter $\alpha$ (dashed, dotted and dash-dotted curves) compared with experimental measurements by different detectors. $\alpha$ has the same interpretation as $\eta$ in Equation~(\ref{eq:numuMDR}). Figure reprinted from~\cite{Cowsik:2012qm}. Copyright 2022 by the American Physical Society.}
	\label{fig:atm_nu}
\end{figure}

Calculations on particle shower development assuming LIV effects can of course still be done to evaluate what signatures experimentalists should be aware of and at what strength level they should appear, given current constraints on LIV. But an atmospheric particle shower is a complex system and if individual LIV parameters are introduced, for example, as a modified dispersion relation as in Equation~(\ref{eq:generalMDR}) for each particle type, the amount of new parameters makes it difficult to make concrete predictions. So simplifications are usually done where LIV is introduced for some particle type but not for others. For example if one focuses on muon production in cosmic ray showers, LIV processes as $\mu \rightarrow e + \gamma$ or $\nu_\mu \rightarrow \nu_e + \gamma$ will change the muon density of the shower at ground level. Additionally, the injection of $e$'s and $\gamma$'s during the development of the shower due to the above processes will induce electromagnetic cascades within the main shower. This anomalous electromagnetic component of the shower can be detected by fluorescence detectors and it will also change  the electron density at ground level. The effect will be more pronounced for inclined showers due to the longer shower development. The effect of LIV on the muon decay probability can be parameterized as 
\begin{equation}
  \Gamma = \frac{1}{\gamma\tau_0} + \eta \frac{\gamma^3}{\tau_0},
\end{equation}
 where $\tau_0$ is the standard muon lifetime, $\gamma$ is the usual Lorentz factor and $\eta$ is the parameter describing the strength of LIV~\cite{Cowsik:1998hp}. A measurement of the electron and muon composition of cosmic ray showers at ground level can thus be used to set a limit on $\eta$. Even with the relatively limited data on inclined  air showers existing at the time, the authors in~\cite{Cowsik:1998hp} could set a limit on $\eta < 10^{-25}$, which was competitive at the time with limits derived from atomic physics or the absence of a Greisen--Zatsepin--Kuzmin (GZK) cutoff (see below).  A similar argument can be developed considering the effect of LIV on atmospheric neutrinos, which distorts the standard pion decay kinematics and therefore the atmospheric neutrino spectrum. When considering the measured atmospheric muon-neutrino spectrum, the dispersion relation (Equation~\ref{eq:generalMDR}) can be simplified to
\begin{equation}
     E^2=m^2+p^2(1+\eta),\label{eq:numuMDR}
\end{equation}
where now $\eta$ is the parameter describing the strength of LIV, and is not limited to values $\pm1$.
Using this probe, a limit on the strength of LIV effects can be set at a level of $10^{-13}$, see Figure~\ref{fig:atm_nu}, although under certain model  assumptions~\cite{Cowsik:2012qm}. 

\begin{figure}[t]
	\begin{minipage}[c]{0.45\textwidth}
		\includegraphics[width=\textwidth]{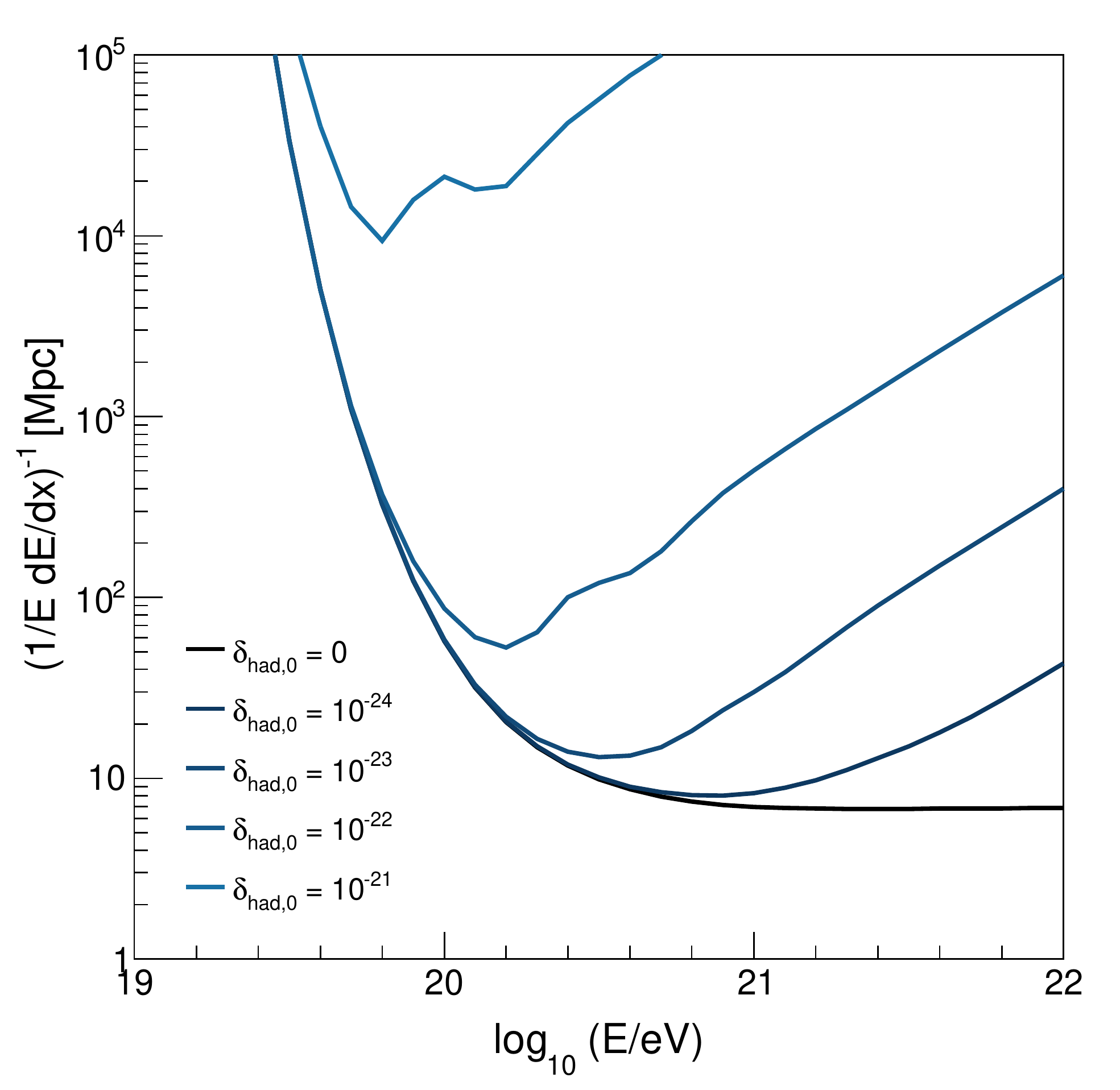}
 	\caption{Attenuation length of cosmic rays due to  photo-pion production as a function of energy for different LIV
	coefficients. $\delta_{\mathrm{had},0}=0$ represents the standard non-LIV case, while stronger LIV effects make the universe more transparent for high energy cosmic rays, challenging the standard interpretation of the GZK cutoff. Figure from~\cite{PierreAuger:2021tog}, reproduced under the CC BY 4.0 license}
    \label{fig:auger_attenuation}
	\end{minipage}\hfill
	\begin{minipage}[c]{0.50\textwidth}
		\includegraphics[height=0.7\textwidth]{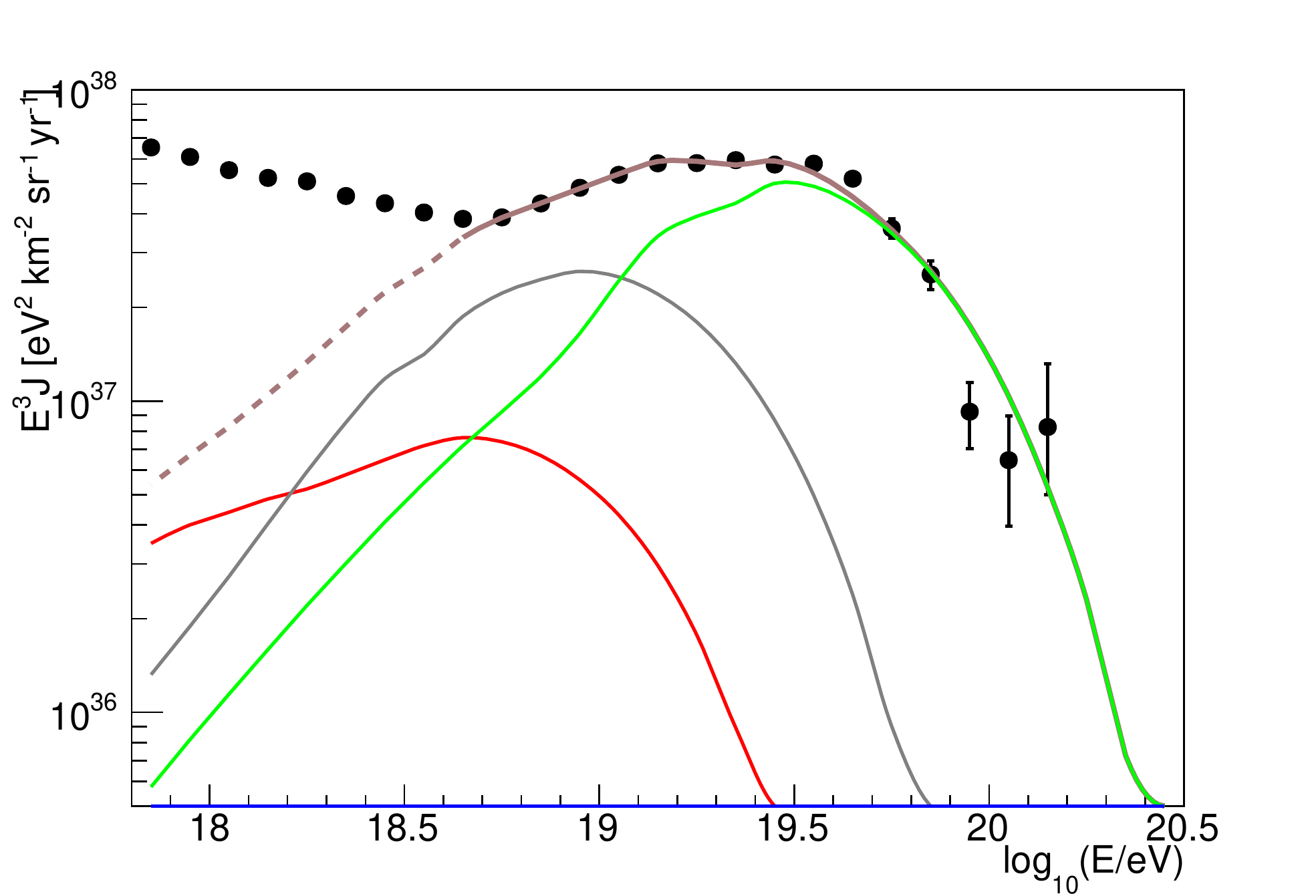}
 	\caption{Energy spectrum for standard (non-LIV) cosmic ray propagation (brown curve) compared to the Pierre Auger Observatory data (black dots). Different colors denote the contribution from different mass numbers (A=1, 2<A<4 and 5<A<22 in red, grey and green, respectively). Figure from~\cite{PierreAuger:2021tog}, reproduced under the CC BY 4.0 license}
    \label{fig:auger_spectrum}
	\end{minipage}
\end{figure}

One of the predicted distinctive features of the cosmic ray energy spectrum is the existence of a cutoff at an energy above $E_{\mathrm{GZK}}\approx 6 \times 10^{19}$~eV due to the interaction of protons with the cosmic microwave background and the extra-galactic background light. Although protons are rare with respect to heavier nuclei at ultrahigh energies, let us consider a proton with an  energy larger than $E_{\mathrm{GZK}}$. In that case, the  process $p + \gamma_{_\mathrm{CMB}} \rightarrow \Delta_{(1232)} \rightarrow p + \pi^{0}$ (or $ \rightarrow n + \pi^{+}$ with the subsequent decay of the neutron into $p + e^{-} +\bar{\nu}_{e}$), will result in a proton of a lower energy, where $\gamma_{_\mathrm{CMB}}$ is a photon from the all-permeating cosmic microwave background with a mean energy in the current epoch of $6 \times 10^{-4}$~eV. The above process has an energy threshold of $E_{\mathrm{GZK}} \geq (m_{\Delta}^2 - m_p^2 )/ 2E_{\gamma{_\mathrm{CMB}}}^2$, so a proton with $E > E_{\mathrm{GZK}}$ will rapidly lose energy through pion photoproduction until its energy falls below $E_{\mathrm{GZK}}$. A similar effect happens for heavier nuclei, which are broken into lighter nuclei of lower energy by photo-dissociation. The GZK cutoff is therefore a universal limitation for the propagation of cosmic rays. It means that if a cosmic ray with an energy $E > E_{\mathrm{GZK}}$ is detected, it either comes from quite close in the universe (the mean free path for a proton with $E > E_{\mathrm{GZK}}$ is about 10 Mpc, quite independent of the initial energy), or some non-standard physics like LIV is at play in its propagation. Turning the argument around, the observation of the GZK cutoff~\cite{HiRes:2007lra,PierreAuger:2008rol} sets a limit on the strength of LIV effects. There is one caveat in this argument and it is that the position of the GZK cutoff we observe is the entangled effect of production at the source and propagation through cosmic distances. If LIV effects become important above a given energy that is achieved during the acceleration or production of the cosmic rays, as well as during propagation, we do not have a way to disentangle these two sources of LIV. Usually, though, it is effects during propagation, through the assumption of non-standard particle dispersion relations or anomalous thresholds for photo-production, that are used to probe LIV with cosmic rays~\cite{Coleman:1997xq,Mattingly:2005re,Martinez-Huerta:2017ulw}. Figure~\ref{fig:auger_attenuation} illustrates the effect that different strengths of LIV (parameterized by the coefficients $\delta$, see Equation~(\ref{eq:dispersion2}) below) have on the attenuation length of cosmic rays. Simulations of the propagation of ultrahigh energy cosmic rays including LIV show that they will interact less with the background photons and they can propagate further than they would do in the absence of LIV. This is the basis for using the observed cosmic ray spectrum to set limits on the $\delta$ parameters. Given that LIV effects should appear at high energies (we do not have any evidence for LIV at the energies reached at accelerators), the modifying term in Equation~(\ref{eq:generalMDR}) is necessarily small, and redefining 
\begin{equation}
\delta_{i,n}\,=\,\frac{\eta_{i,n}}{E_\mathrm{QG}^n}
\label{eq:delta}
\end{equation}
 where $n$ represents the order of LIV and $i$ denotes a particle species, Equation~(\ref{eq:generalMDR}) can be recast as a power expansion in energy,
\begin{equation}
 E_i^2\,=\,p_i^2\,+\,m_i^2\,+\,\Sigma_{n=0}^{N} \delta_{i,n}E_i^{2+n}
 \label{eq:dispersion2}
\end{equation}

Measurements of the energy spectrum and cosmic ray flux composition at ultrahigh energies by the Pierre Auger Observatory, see Figure~\ref{fig:auger_spectrum}, leave little room for LIV effects in the propagation of ultrahigh energy cosmic rays and allow to set very stringent limits on several $\delta$ parameters for hadrons: $\delta_{\mathrm{had},0} < 10^{-19}$, $\delta_{\mathrm{had},1} < 10^{-38}~eV^{-1}$ and $\delta_{\mathrm{had},2} < 10^{-57}~eV^{-2}$ at  5$\sigma$ confidence level~\cite{PierreAuger:2021tog}.  The difficulty in this kind of analysis is the modeling of the ``standard'' propagation scenario against which data are compared. On the astrophysical side, the distribution of the diffuse extra-galactic background light is difficult to asses due to the large local backgrounds. On the particle physics side, photo-nuclear cross sections need to be extrapolated over several orders of magnitude in energy with respect to measured values. These unknowns add uncertainties to the simulations of extensive air showers and therefore on the extracted mass composition through the measured $X_{max}$, the depth at which the shower contains the maximum number of particles. Meaningful limits on LIV coefficients can still be set as long as the characteristics of the observed cosmic ray spectrum and composition are compatible with expectations within uncertainties. But a detailed  understanding of the standard model physics in the propagation and interaction of ultrahigh energy cosmic rays is  needed if ever a deviation from expectations is to be explained by new physics.

 There is another process linked to the existence of LIV that can also lead to a suppression of ultrahigh energy cosmic rays: ``gravitational Cherenkov radiation'', an effect that has been argued that could take place when the speed of a particle exceeds the speed of propagation of gravity. In the presence of LIV the radiation of a graviton by cosmic rays with speeds larger than the speed of gravitational waves would result in an unconventional energy loss of the particle, which could become observable over cosmological travel distances as a lack of cosmic rays above certain energy. Therefore, the observation of ultrahigh energy cosmic rays can be used to set stringent limits on the strength of LIV. Using the SME as a benchmark, the authors in~\cite{Kostelecky:2015dpa} arrive at an expression for the limit on the coefficients $s^d$responsible for LIV as
 \begin{equation}
s^d(\hat{p}) \, < \, \sqrt{  \frac{{\cal{F}}(d)}{G_N E^{2d-5} L}  }
 \label{eq:gravitational_cherenkov}
 \end{equation}
  where $d$ is the mass dimension of the corresponding operator in the SME, $\cal{F}$ is a dimensionless numerical factor that depends on $d$ but also on the particle species under consideration, $G_N$ is Newton's constant and $E$ and $L$ are the energy and travel distance of the cosmic ray which arrives at Earth from the direction $\hat{p}$. 
  Since LIV effects can depend on the direction of arrival of cosmic rays, the $s$ coefficients further depend on orientation in a solar-centered coordinate system through the quantum number $j$ when expressed as an expansion in spherical harmonics, $s^d(\hat{p})=\sum{Y_{j,m}(\hat{p}) s^d_{j,m}}$. 
   Assuming a value of the order of Mpc for L, typical distance to the closest AGN,  constraints on $s^d$ for different dimensions (order of $s^d < 10^{-14}, 10^{-30}$ GeV$^{-2}$, $10^{-46}$ GeV$^{-4}$ for $d=4,6,8$ respectively, slightly dependent on orientation through the quantum number $j$) have been set from the measured energies of cosmic rays~\cite{Kostelecky:2015dpa}. These are quite competitive constraints from a quite simple analysis, which shows again the advantage of using cosmic messengers to probe tiny new physics effects.
  
\subsection{Gravitational waves}
\label{subsec:GWs}
September 14th, 2015 marked the beginning of gravitational-wave astronomy with the direct detection of the event GW150914~\cite{LIGOScientific:2016aoc} by the LIGO~\cite{LIGO:01} and Virgo~\cite{Virgo:01, Virgo:2022fxr} collaborations, an event compatible with the signal predicted by general relativity for the spiral and merger of two black holes of 36 and 29 solar masses. Since then, several black hole and neutron star mergers have been detected. Gravitational waves, along with cosmic rays and neutrinos, complete our non-electromagnetic probes of the universe and, as those other messengers, also provide a glimpse into fundamental physics. In general relativity, gravitational waves propagate with the speed of light and this provides a reference to test LIV effects either in the gravitational sector alone (if the gravitational wave speed depends on frequency and/or arrival direction) or in relation to the propagation of other messengers (using the collapse of binary systems to provide a clock for the event and comparing arrival times with other messengers as far as their emission can be assumed to be simultaneous). These are the same kind of tests that we have mentioned with other messengers above, but provide a useful complementary probe into new physics at the Planck scale.

If we concentrate on the specific tests that can be done with information from the propagation of gravitational waves alone\footnote{Tests using relative timing with respect to other messengers or direction-dependent observables are similar to what was discussed in section~\ref{subsec:NUs}}, the handles that can be used are the deformation of the waveform due to different propagation speeds for different frequencies (dispersion) and changes in the polarisation modes (birefringence) with respect to the expected signal without LIV. The SME can also be used here as a generic tool to test LIV effects in an effective way by writing the Lagrangian as an  expansion of terms of mass dimension $d$ in a 
linearized metric, $g_{\mu\nu}=\eta_{\mu\nu}+h_{\mu\nu}$, where $h_{\mu\nu}$ is understood as a perturbation to the flat Minkowski metric $\eta_{\mu\nu}$~\cite{Mewes:2019dhj}. In the case of strong dispersion or birefringence effects the gravitational wave event might be ``diluted'' and fall under the detector threshold, not producing the typical chirp signature. As with other messengers, multiple observations from different directions in the sky are necessary to study LIV effects with gravitational waves since Lorentz violation can depend on direction: the above mentioned effect can render observations of gravitational waves from a given direction difficult if indeed LIV is at play.

\begin{figure}[t]
	\includegraphics[width=.66\textwidth]{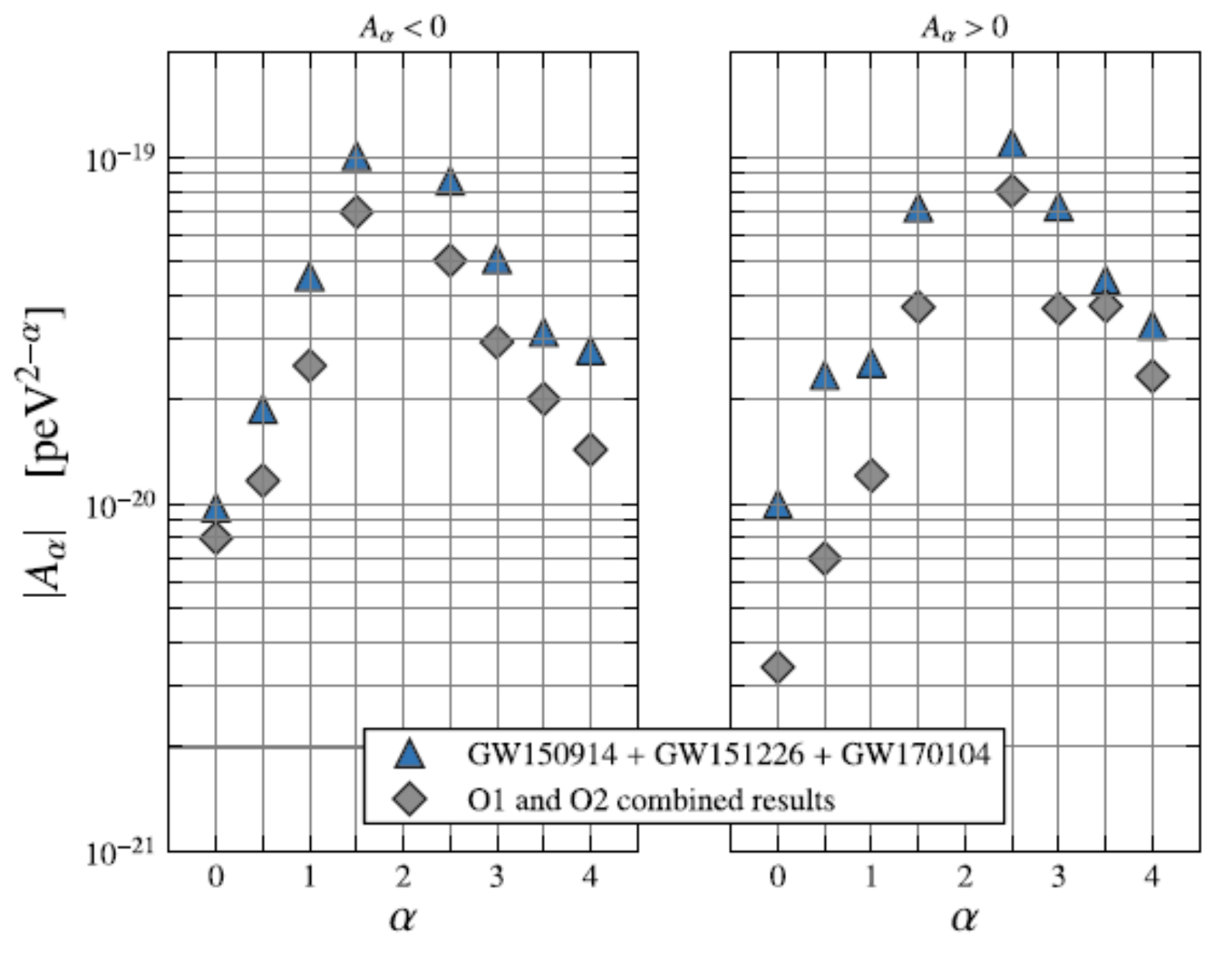}
	\caption{90\% upper limits on the parameter $A_{\alpha}$. Blue triangles denote the results obtained on GW150914, GW151226, and GW170104, which are updated with respect to the results previously published in~\cite{LIGOScientific:2017bnn}. Grey diamonds represent results obtained jointly on GW150914, GW151012, GW151226, GW170104, GW170608, GW170729, GW170809, GW170814, GW170818, and GW170823. The scale of PeV was chosen because it is equivalent to 250\,Hz, which is close to the frequencies in which gravitational wave detectors LIGO and Virgo are the most sensitive. Figure reprinted from~\cite{LIGOScientific:2019fpa}. Copyright 2022 by the American Physical Society.}
	\label{fig:WG_limits}
\end{figure}

The event GW150914 alone was already used in~\cite{Kostelecky:2016kfm} to set limits on LIV coefficients with dimension 4 and 5 from the absence of a splitting in the signal polarisation modes. This sets constraints in the dispersion relation of gravitational waves when written as a function of Lorentz violating coefficients. However the strongest limits on modifications of the propagation of gravitational waves currently come from analyses of the LIGO/Virgo collaborations themselves using the events collected in the GWTC-1 catalog~\cite{LIGOScientific:2018mvr,LIGOScientific:2019fpa}. Taking a pure phenomenological approach, a Lorentz-violating dispersion relation can be parameterized as
\begin{equation}
E^2=p^2c^2 + A_{\alpha}p^{\alpha}c^{\alpha}.    
\end{equation}
Limits on the coefficients $A_{\alpha}$ can be set by comparing the characteristics of the detected signals with the predicted waveforms obtained from general relativity (see Figure~\ref{fig:WG_limits}). 

As just argued, cataclysmic events produce gravitational waves over a short period of time that provide a timing for the emission that, in conjunction with other messengers, can be used to probe general relativity and deviations thereof. But gravitational waves should also be emitted by non-spherically symmetric rotating neutron stars (pulsars). Due to the strong gravity on their surface, it is difficult for a neutron star to maintain mass distribution irregularities or asymmetric deformations that would make it non spherical. But the distribution of the magnetic field inside the star, internal convection or accretion from a close companion can cause a small ellipticity $\epsilon$ on rotating neutron stars of the order of $\epsilon <  10^{-8}-10^{-6}$~\cite{Lasky:2015uia}.  Since pulsars can be found relatively close and well spatially resolved, a detection of continuous, practically monochromatic gravitational waves from a pulsar would provide a unique additional way to test general relativity, but deviations of it as well, i.e. LIV effects~\cite{Zimmermann:1980ba,Xu:2020zxs}. The existence of a preferred direction in space, a source of LIV, can modify the rotation of free neutron stars by producing a torque that forces the angular momentum of the star to precess around an axis aligned with the spatial preferred direction. This process results in the continuous emission of gravitational waves with a different spectrum than those emitted by non-spheroidal stars just rotating in isotropic space. Searches for continuous emission of gravitational waves from rotating pulsars have been carried out, although without success so far~\cite{LIGOScientific:2017hal} since the expected frequency of the gravitational waves from rotating neutron stars (about 1~Hz) lies near the lower sensitivity limit of current detectors.

 Third-generation gravitational wave detectors currently in R\&D, like the Einstein Telescope~\cite{DiPace:2022uzn}, will be able to increase the detection sensitivity to frequencies as low as 1~Hz (the current lower threshold of LIGO is about 10~Hz) while space-based interferometers like LISA~\cite{Amaro-Seoane:2012vvq} will have a peak sensitivity at $10^{-2}$~Hz, sufficient to explore gravitational wave emission from binaries in our Galaxy and vastly increase the sensitivity to deviations from general relativity~\cite{Perkins:2020tra}.

\section{Outlook}
\label{sec:outlook}

Cosmic searches for Lorentz invariance violation are an attempt at probing effects of quantum gravity. The formulation of a quantum theory of gravity proved to be quite a challenge so far. An experimental detection of a possible effect of quantum gravity, hinting at what the quantum nature of gravity might be, could be decisive for reaching this paramount scientific goal. However, quantum gravity is expected to manifest at energies on the order of Planck scale, far beyond the reach of accelerator experiments. That is where experiments with cosmic messengers come into play. The energies of astrophysical gamma rays, neutrinos, and cosmic rays can be orders of magnitude higher than energies attainable in accelerators, and more likely to probe quantum gravity. In addition, these messengers traverse enormous distances, over which tiny effects of quantum gravity are expected to accumulate enough to be detected. 

The highest measured energies of cosmic messengers are still orders of magnitude below the Planck scale, so what is really tested are effective theories which emulate consequences of quantum gravity at lower energies. LIV as one possible effect of quantum gravity can be introduced by adding terms in the Standard Model Extension, while modifying dispersion relation as in Equation~(\ref{eq:generalMDR}) is a very simple way of modelling physics outside of Standard Model. 
Modifying the dispersion relations of particles leads to a plethora of possible new effects. Some of these, such as energy-dependent group velocity of massless particles, photon instability, vacuum bremsstrahlung or Cherenkov radiation, vacuum birefringence, etc. are forbidden in the standard special-relativistic scenario. On the other hand, propagation speed of massive particles, particle interaction thresholds (e.g. gamma-ray absorption on the background photon fields, GZK cutoff), neutrino flavour oscillations, etc. are modified with respect to the same phenomena predicted by the standard physics. 

Detectors of cosmic messengers already yielded some important results in the searches for traces of Lorentz invariance violation. So far, no such effects were found. However, strong constraints have been set on some of LIV parameters. The energy scale at which LIV effects could manifest for some of them has been bounded to above the Planck energy; in some cases, even several orders of magnitude above. A census of experimental tests and limits on LIV can be found in the QG-MM Catalogue \cite{QGMMCatalogue}\footnote{\href{https://qg-mm.unizar.es/wiki}{https://qg-mm.unizar.es/wiki/}}, created and maintained by the COST Action 18108.

Experimental studies will surely become more sensitive and keep probing new regions of parameter space,  especially with the introduction of new experiments such as CTA~\cite{CTA:2020hii}, IceCube-Gen2, KM3NET and the third generation of gravitational wave detectors. 
However, the vast majority of these test were performed using single messenger data sets. In fact, only a handful astrophysical events were confirmed as multi-messenger until now. 
With more sensitive instruments, we expect to detect multi-messenger events at a greater rate, enabling more frequent and more sensitive multi-messenger tests of LIV, leading us closer to be able to paint the whole picture of what quantum gravity might have in stock.

\section{Further reading:}
It is impossible in such a short review to give credit to all the relevant work published in the subject by theorists and experimentalists. Since we have tried to keep the references to the point, we list here a few additional papers that, even without the aim of being comprehensive, can be useful for a reader interested in the subject of this chapter:\par

\begin{itemize}
    \item The COST Action CA18108 "Quantum gravity phenomenology in the multi-messenger approach" has published a comprehensive review of theory, phenomenology, and experimental searches for effects of quantum gravity in astrophysical observations ``Quantum gravity phenomenology at the dawn of the multi-messenger era --- A review'' \cite{Addazi:2021xuf}.
    \item A detailed description of LIV studies performed with Cherenkov telescopes can be found in \cite{Terzic:2021rlx} and references therein. Various effects, analysis methods, and results are mutually compared and discussed.
    \item We already mentioned some these, but it does not hurt to repeat: there are excellent reviews on the phenomenology and signatures of LIV in high energy neutrinos~\cite{Christian:2004xb,Roberts:2021vsi,Torri:2020dec,Stecker:2017gdy,Diaz:2014yva,Stecker:2014oxa,Hooper:2005jp,Stecker:2022tzd}.
    \item The ingredients needed for searches for LIV with cosmic rays are well covered in~\cite{Bietenholz:2008ni},
    \item The search for fundamental physics with gravitational waves is covered in~\cite{Will:2014kxa},\cite{Perkins:2020tra} and \cite{Mewes:2019dhj}

\end{itemize}

\subsection{Problems}
\label{sec:LIVexperimentalProblems}
\begin{enumerate}
    \item \label{problem:TimeDelay} Starting from the dispersion relation for photons (Equation~\ref{eq:generalMDR}), calculate modified photon group velocity (Equation~\ref{eq:photonvelocity}). Derive the expression for the arrival time delay between two photons emitted at the same time.\\ Hint: follow the procedure from \cite{Jacob:2008bw}.
    \item \label{problem:TimeDelaySim} Generate a data sample of 1000 gamma rays. Each event should be characterised with emission time $t$, given in seconds, and energy $E$, given in GeV. Let the distribution of emission times follow normal distribution, and distribution of energies follow a power law $\Phi(E)\propto E^{-\gamma}$, with $\gamma = 2.5$. Use Equation~(\ref{eq:t_prime}) to calculate the detection time for each gamma ray assuming they were emitted from a source at redshift 0.5 and $E_\mathrm{QG} = E_\mathrm{Pl} = 1.22\times10^{19}$\,GeV. Plot the distribution of arrival times and compare it to the distribution of the emission times. Do the same both for $n=1$ and $n=2$. Repeat the exercise for different values of $E_\mathrm{QG}$ and different redshifts.\\ Note that the same exercise could be performed for neutrinos, as well as cosmic rays. What is the problem when using cosmic rays for such analysis?
    \item \label{problem:BWcrosssection} Work out the reaction threshold and the invariant mass for the Breit--Wheeler process in standard special relativity, and in LIV. Compare your results to equations (\ref{eq:modified_energy_threshold}) and (\ref{eq:modified_center_of_mass}).
    \item \label{problem:neutrinoOscillations} Derive the neutrino flavour transition probability for a two-flavour scenario (see Equation~\ref{eq:std_osc}). Then repeat the derivation assuming modified neutrino dispersion relation to obtain Equation~\ref{eq:osc_prob2}.\\ Hint: follow the procedure from \cite{Morgan:2008fof}.
    \item \label{problem:inertialSystems} A cosmic ray (proton) propagates towards Earth with the kinetic energy $T$ and scatters on a stationary proton in the atmosphere. Calculate the kinetic energies of these two protons in the center of mass system. What is the cosmic ray energy that would be equivalent to proton--proton collision at the LHC at 14\,TeV?
    \item \label{problem:GZKcutoff} Calculate the threshold energy for the GZK cutoff.
\end{enumerate}

\bibliographystyle{ieeetr}
\bibliography{main.bib}

\end{document}